\documentclass{mn2e}
\usepackage{mathtools}
\usepackage{xcolor}
\usepackage{graphicx}
\usepackage{enumerate}
\usepackage{amsmath}
\usepackage{breqn}
\usepackage{epstopdf}

\usepackage{array}
\usepackage{amssymb}
\usepackage{caption}
\usepackage{multirow}
\let\bibhang\relax
\usepackage{natbib}
\usepackage{array}
\usepackage{hyperref}
\graphicspath{{./pics/}}
\begin{document}
\title[Simulation tests of chameleon gravity]{Simulation tests of galaxy cluster constraints on chameleon gravity}
\author[Wilcox, Nichol, Zhao, Bacon, Koyama, Romer]
{Harry Wilcox,$^1$ Robert C. Nichol,$^1$ Gong-Bo Zhao,$^1$ David Bacon,$^1$ \newauthor Kazuya Koyama,$^1$ A. Kathy Romer$^2$\\
$^1$ Institute of Cosmology and Gravitation, University of Portsmouth, Dennis Sciama Building, Portsmouth, PO1 3FX, UK\\
$^2$ Dept. of Physics and Astronomy, University of Sussex, Falmer, Brighton, BN1 9QH
}
\maketitle
\begin{abstract}
We use two new hydrodynamical simulations of $\Lambda$CDM and $f(R)$ gravity to test the methodology used by \citealt{2015MNRAS.452.1171W} (W15) in constraining the effects of a fifth force on the profiles of clusters of galaxies. We construct realistic simulated stacked weak lensing and X-ray surface brightness cluster profiles from these cosmological simulations, and then use these data projected along various lines-of-sight to test the spherical symmetry of our stacking procedure. We also test the applicability of the NFW profile to model weak lensing profiles of clusters in $f(R)$ gravity. Finally, we test the validity of the analytical model developed in W15 against the simulated profiles. Overall, we find our methodology is robust and broadly agrees with these simulated data. We also apply our full Markov Chain Monte Carlo (MCMC) analysis from W15 to our simulated X-ray and lensing profiles, providing consistent constraints on the modified gravity parameters as obtained from the real cluster data, e.g. for our $\Lambda$CDM simulation we obtain $|f_{\rm{R}0}| < 8.3 \times 10^{-5}$ (95\% CL), which is in good agreement with the W15 measurement of $|f_{\rm{R}0}|  < 6 \times 10^{-5}$. Overall, these tests confirm the power of our methodology which can now be applied to larger cluster samples available with the next generation surveys.  

\end{abstract}
\section{Introduction}

Over the last century, General Relativity (GR) has provided a robust theoretical framework for cosmology. However, a major challenge for this framework is the discovery of acceleration in the late-time expansion history of the Universe (\citealt{1998AJ....116.1009R};
\citealt{1999ApJ...517..565P}), leading to the need for a cosmological constant in the Friedmann equations, and/or a possible modification to GR on cosmological scales (\citealt{2012PhR...513....1C}, \citealt{2015arXiv150404623K}).

Modified theories of gravity typically require an additional scalar field which couples to matter, giving rise to a fifth force \citep{2013ApJ...779...39J}.  GR has been well tested on terrestrial and Solar System scales \citep{2012CQGra..29r4002W}, so if such an additional force did exist, it must be suppressed, or ``screened'', in dense environments to avoid detection.  

A popular method for screening such a field is the chameleon mechanism \citep{Khoury:2003aq}.  In this theory, the coupling strength of the additional scalar field is dependent on the local gravitational potential. In regions with deep gravitational potential wells, the field is screened and gravity behaves like GR, while in areas of shallow gravitational potentials, gravity becomes modified and deviates from GR \citep{2014AnP...526..259L}.

The chameleon fifth force obeys

\begin{equation}
F_{\phi} = -\frac{\beta}{M_{\rm{Pl}}} \nabla \phi,
\end{equation}

\noindent where $M_{\rm{Pl}}$ is the Planck mass, $\phi$ is the scalar field and $\beta$ is the scalar field coupling strength to matter.  The value of the scalar field on the cosmological background ($\phi_{\infty}$) measures the efficiency of the screening \citep{2012PhRvD..86j3503T}.  The chameleon mechanism is then characterised by the parameters $\beta$ and $\phi_{\infty}$.

There is a particular set of chameleon gravity models, known as $f(R)$ gravity, where $\beta = \sqrt{1/6}$ \citep{2007PhRvD..75l4014C}.  In these models the fifth force is mediated by an additional degree of freedom, namely $f_{\rm{R}}=\text{d}f/ \text{d}R$, which at $z=0$ has a value of $|f_{\rm{R}0}|$ \citep{2010RvMP...82..451S}.  We can relate $f_{\rm{R}}$ to the screening efficiency ($\phi_{\infty}$ above) by {\citep{2014arXiv1407.0059J}

\begin{equation}
f_{\rm{R}}(z) = -\sqrt{\frac{2}{3}} \frac{\phi_{\infty}}{M_{\rm{Pl}}}.
\label{eq:f(R)}
\end{equation}

In our previous paper (\citealt{2015MNRAS.452.1171W}, or W15 hereafter), we investigated a possible observational signature of chameleon gravity using clusters of galaxies as first studied by \cite{2014JCAP...04..013T}. These works postulate that the additional fifth force could be unscreened in the outskirts of groups and clusters of galaxies, leading to an observed difference between the X-ray and gravitational weak lensing profiles of the clusters. Therefore, constraints on the modified gravity parameters can be obtained by comparing measurements of both the X-ray and weak lensing cluster profiles for a set of clusters. For chameleon gravity, cluster observations provide the most stringent constraints to date on cosmological scales (mega to gigaparsec scales) and are complementary to the solar system and dwarf galaxy tests which probe smaller scales (up to kiloparsec scales).

\citet{2014JCAP...04..013T} achieved a constraint of $|f_{\rm{R}0}| < 6 \times 10^{-5}$ from observations of just the Coma Cluster. In W15, we applied the same technique to a stacked profile of 58 X-ray clusters taken from the XMM Cluster Survey (XCS; \citealt{1999astro.ph.11499R}) with weak lensing data from the Canada France Hawaii Telescope Lensing Survey (CFHTLenS; \citealt{2012MNRAS.427..146H}). Using a multi-parameter Monte Carlo Markov Chain (MCMC) analysis, we constrained the values of the two chameleon gravity parameters ($\beta$ and $\phi_{\infty}$), finding the data was consistent with GR, i.e., we did not require a fifth force. In the case of $f(R)$ gravity ($\beta = \sqrt{1/6}$), we constrained $|f_{\rm{R}0}| < 6 \times 10^{-5}$ (95\% confidence), which is similar to the Coma cluster limit. These constraints remain some of the strongest constraints on the background field amplitude on cosmological scales (for a review see \cite{2014AnP...526..259L} and references therein).

The work of W15 made several simplifying assumptions, including: {\it i)} All clusters were in hydrostatic equilibrium, with no significant additional non-thermal pressure affecting their profiles; {\it ii)} Stacking clusters produces a fair representation of spherically-symmetrical profiles by minimising line-of-sight projection effects; {\it iii)} Dark matter haloes are well described by an NFW profile in chameleon gravity \citep{1996ApJ...462..563N}.

The first of these assumptions was tested in W15 where we found that our cluster profiles were consistent with no additional (non-thermal) pressure at, and beyond, the virial radius of the stacked cluster profiles. The other assumptions are tested in this paper, which also provides a confirmation of the analytical modelling presented in W15 for simultaneously describing changes to the X-ray and lensing profiles of clusters due to modifications of gravity. 

We achieve these tests using two new hydrodynamical cosmological simulations; one evolved using the concordance $\Lambda$CDM+GR model, and the other evolved using $f(R)$ gravity with a background field amplitude of $|f_{\rm{R}0}| = 10^{-5}$. This value of $|f_{\rm{R}0}|$ was chosen to be consistent with present observational limits on this parameter, as smaller values would have resulted in a modification to gravity that could not be measured within these present simulations (due to an insufficient number of haloes), nor with the data in W15.  

In Section 2, we describe the cosmological simulations used throughout this paper and the techniques used to generate suitable simulated data products. In Section 3, we discuss the creation of the stacked X-ray and weak lensing cluster profiles, and test the assumptions in W15 discussed above. In Section 4, we present results from our MCMC fitting of the simulated stacked cluster profiles. We discuss these results in Section 5. 

\begin{figure*}
\centering
\begin{minipage}{0.49\textwidth}
\centering
%\exedout % first figure itself
\includegraphics[width=\textwidth]{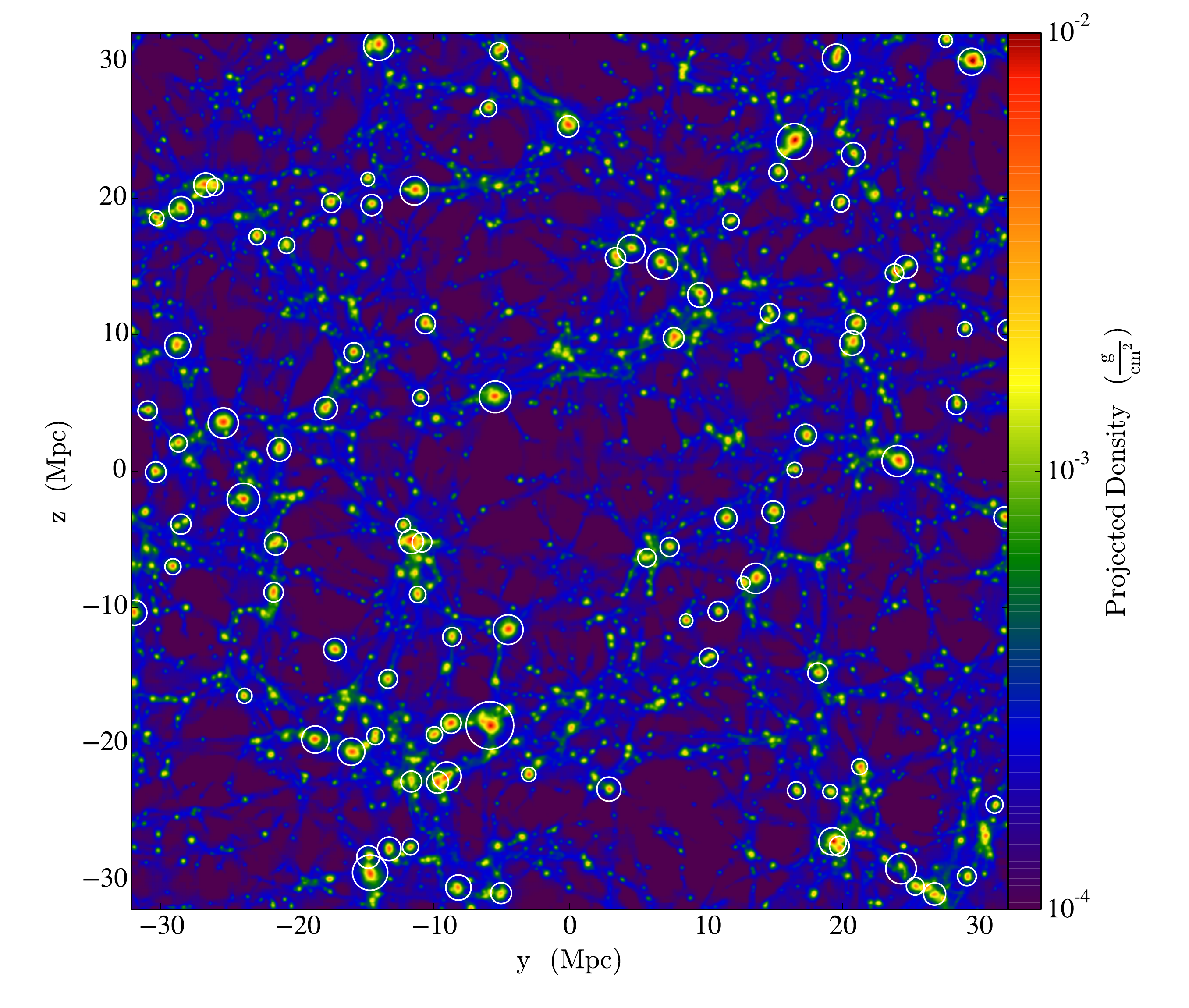}
\caption{Snapshot of the total mass density for the $\Lambda$CDM+GR simulation at $z=0.4$. We highlight all halos above $\textrm{M}>10^{13}\textrm{M}_{\odot}$. The simulation has been projected along one side of the simulation box (128 Mpc/h cube). }
\label{fig:halos_LCDM}
\end{minipage}\hfill
\begin{minipage}{0.49\textwidth}
\centering
%\exedout % second figure itself
\includegraphics[width=\textwidth]{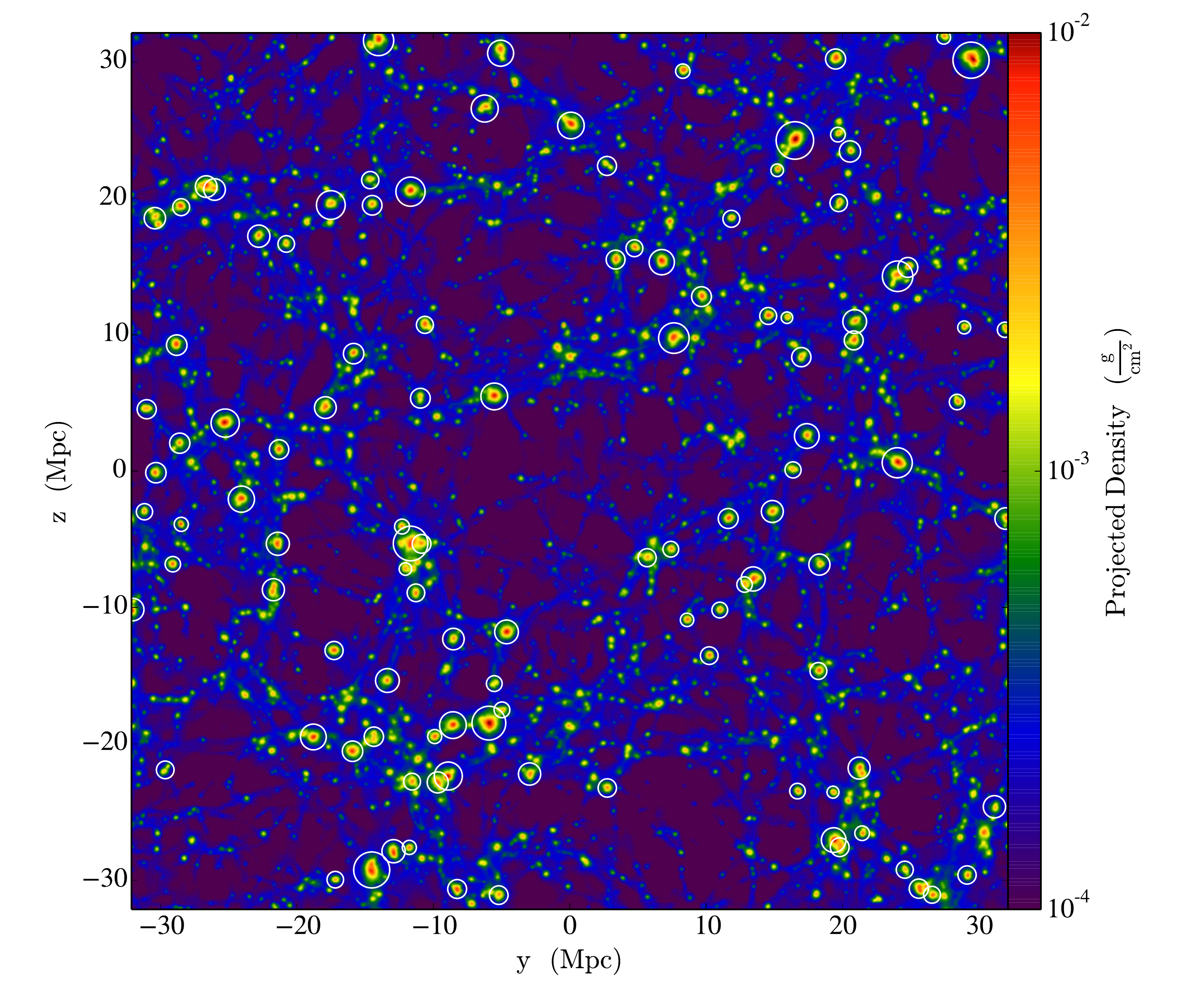}
\caption{The same as Figure \ref{fig:halos_LCDM} but for the $f(R)$ simulation.\textcolor{white}{Some text to give caption second line. More padding text, copy as needed. More padding text, copy as needed. More padding text, copy as needed.}}
\label{fig:halos_MG}
\end{minipage}
\end{figure*}

\section{Simulating Clusters}

\subsection{Cosmological Simulations}
\label{sims}

In this paper we use two new hydrodynamical simulations created using the {\tt MGENZO} software; a variant of the {\tt ENZO}\footnote{Available at http://enzo-project.org/} software, but working with modified gravity theories (see Zhao et al. in prep). We provide an overview of the theoretical details of these simulations in Appendix A.

In particular, we use one available $f(R)$ simulation with $|f_{\rm{R0}}|=10^{-5}$ and $n=1$ (see Equation A1 for reference) and, for comparison, a $\Lambda$CDM+GR simulation. The constraints in this paper can be converted to other values of $n$ as discussed in Terukina et al. (2014). Both simulations have $2\times128^3$ particles of mass $4 \times 10^{11} \rm{M}_{\odot}$ in a cubic box of 128 Mpc/h on a side. Both simulations have identical initial conditions and background cosmological parameters, namely $\Omega_{b}=0.044$, $\Omega_{cdm} = 0.226$, $\Omega_{\Lambda} = 0.73$ and $H_{0}=71 \rm{km s}^{-1} \rm{Mpc}^{-1}$ \citep{planck2013} Each simulation is evolved to $z=0.4$, which is close to the median redshift of the W15 cluster sample (${\bar{z}_{cluster}}=0.33$).

Each simulation provides the location and temperature of all particles (assuming the equipartition of kinetic energy), which are then used to determine the density and pressure of the gas (assuming the ideal gas law). Neither simulation includes any additional feedback processes (e.g. from active galactic nuclei or supernovae) and therefore will not include an additional non-thermal pressure component that could affect the temperature and density of gas in galaxy clusters (\citealt{2004MNRAS.351..237R}, \citealt{2015arXiv151107904O}). 

In Figures \ref{fig:halos_LCDM} and \ref{fig:halos_MG}, we show the projected density (baryons and dark matter) for the two simulations at $z=0.4$.

\subsection{Finding Dark Matter Haloes}
\label{sec:finding_haloes}

To replicate the analysis of W15, we must find clusters in our simulations and stack their profiles. We therefore use the ``Rockstar'' Friends-of-Friends (FOF) algorithm \citep{2013ApJ...762..109B} to locate the main dark matter haloes in both simulations (Section \ref{sims}). We use the default linkage length of $0.28$ times the mean particle separation. For each simulation, we obtained a catalogue of the halo locations (centre-of-mass) and their masses.  All haloes are at $z=0.4$ which again is close to the median of the W15 clusters (${\bar{z}_{cluster}}=0.33$).

We impose a threshold on the halo mass of $\textrm{M}>10^{13}\textrm{M}_{\odot}$. This is a compromise to ensure we have sufficient haloes for our tests, while still containing the more massive halos in the simulations which were closest to the typical cluster masses used in W15. At redshifts below $z=0.4$ (for consistency with W15), the W15 sample has a typical mass of $\simeq 8 \times 10^{13}\textrm{M}_{\odot}$, which is higher than the typical mass of the halos above our threshold ($\simeq 4 \times 10^{13}\textrm{M}_{\odot}$). This is due to a few massive clusters in the W15 sample ($T_x>2.5$ keV) that are not present in the simulations. The modes of the mass (or X-ray temperature) distribution are similar between the real and simulated samples. We return to this point in Section 5 when discussing our results. 
  
Above this threshold mass, we find 103 clusters (or haloes) in the $\Lambda$CDM+GR simulation and 113 clusters in the $f(R)$ simulation. Interestingly, it has been shown before (\citealt{2009PhRvD..79h3518S}, \citealt{2014JCAP...03..021L}, \citealt{winther2015}) that the abundance of massive clusters is enhanced in the presence of a fifth force (with $ |f_{\rm{R}0}| \sim 10^{-5}$), and this signal can be used to constraint modified gravity models (see \citealt{2015PhRvD..92d4009C}). We do not consider this signal further here as our technique focuses on the shape of cluster profiles, not their relative abundance.

We found $\simeq 4\%$ of the volume in our $f(R)$ simulation was contaminated by unrealistic particle velocities, leading to an extremely low density ($<10^{7}{\rm M_{\odot}} / {\rm Mpc}^{-3}$), but exceptionally hot ($>5 \times 10^{8}{\rm K}$), extended bubble surrounding the most massive dark matter halo in the simulation. This bubble is potentially caused by the lack of realistic feedback mechanisms (Section \ref{sims}). The bubble affects 14 nearby clusters, which are enclosed by it, on average doubling their temperature profiles at the virial radius.  Therefore we remove these 14 haloes, leaving 99 clusters in total for the $f(R)$ simulation. This bubble is not visible in the density map in Figure \ref{fig:halos_MG}. n Figures \ref{fig:indv_clust_LCDM} and \ref{fig:indv_clust_MG}, we show four typical (randomly-selected) haloes from both simulations.

To ensure our results are valid regardless of the precise details of our mass threshold (and thus cluster abundance), we repeated our full analysis using only the most massive 50 clusters (halos) in both simulations. In both cases, we exclude the contaminated 14 clusters (halos) discussed above. As the initial conditions are the same in both simulations, these halos represent the same over-densities in both simulations. As expected, the constrains on $\beta$ and $\phi_{\infty}$ are $\sim35$\% weaker than for the  full sample presented in Figures 12 and 13 due to the smaller sample size and the fact that the screening mechanism is more efficient in massive clusters. That said, these additional constraints are similar in shape and scale as the full sample, indicating that the details of our mass threshold do not systematically bias our result and our constraints on $\beta$ and $\phi_{\infty}$ are dominated by statistical uncertainties (i.e. the number of clusters available to us).

In Figures \ref{fig:indv_clust_LCDM} and \ref{fig:indv_clust_MG}, we show four typical (randomly-selected) haloes from both simulations.

\begin{figure*}
\centering
\begin{minipage}{0.49\textwidth}
\centering
%\exedout % first figure itself
\includegraphics[width=\textwidth]{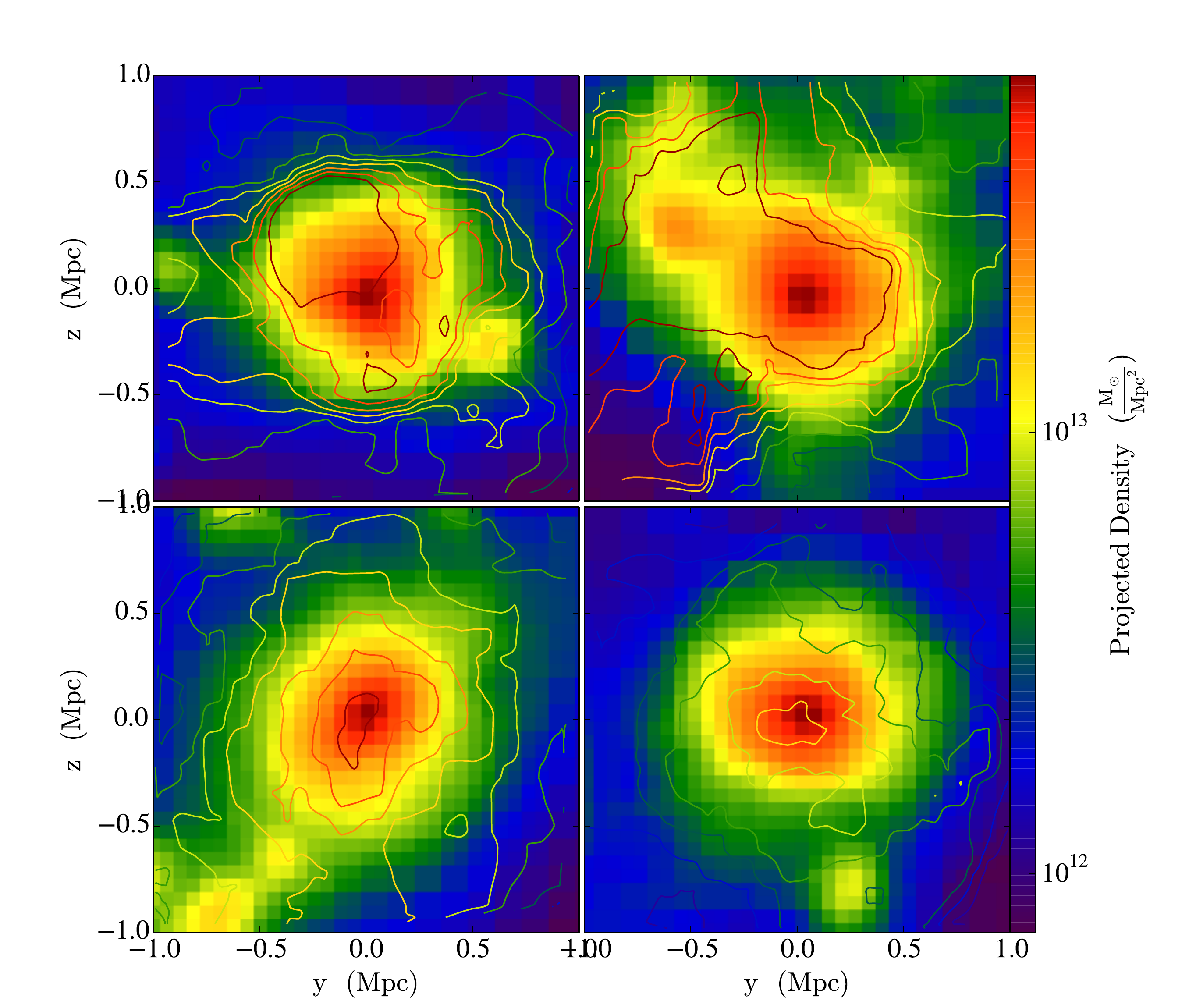}
\caption{A mosaic of four (projected) cluster profiles taken from the $\Lambda$CDM+GR simulation at $z=0.4$. The coloured contours are the gas density, while the black contour lines are the gas temperature. Here the redder contours denote higher temperatures and bluer lower over a scale of 2keV to 0.5keV}
\label{fig:indv_clust_LCDM}
\end{minipage}\hfill
\begin{minipage}{0.49\textwidth}
\centering
%\exedout % second figure itself
\includegraphics[width=\textwidth]{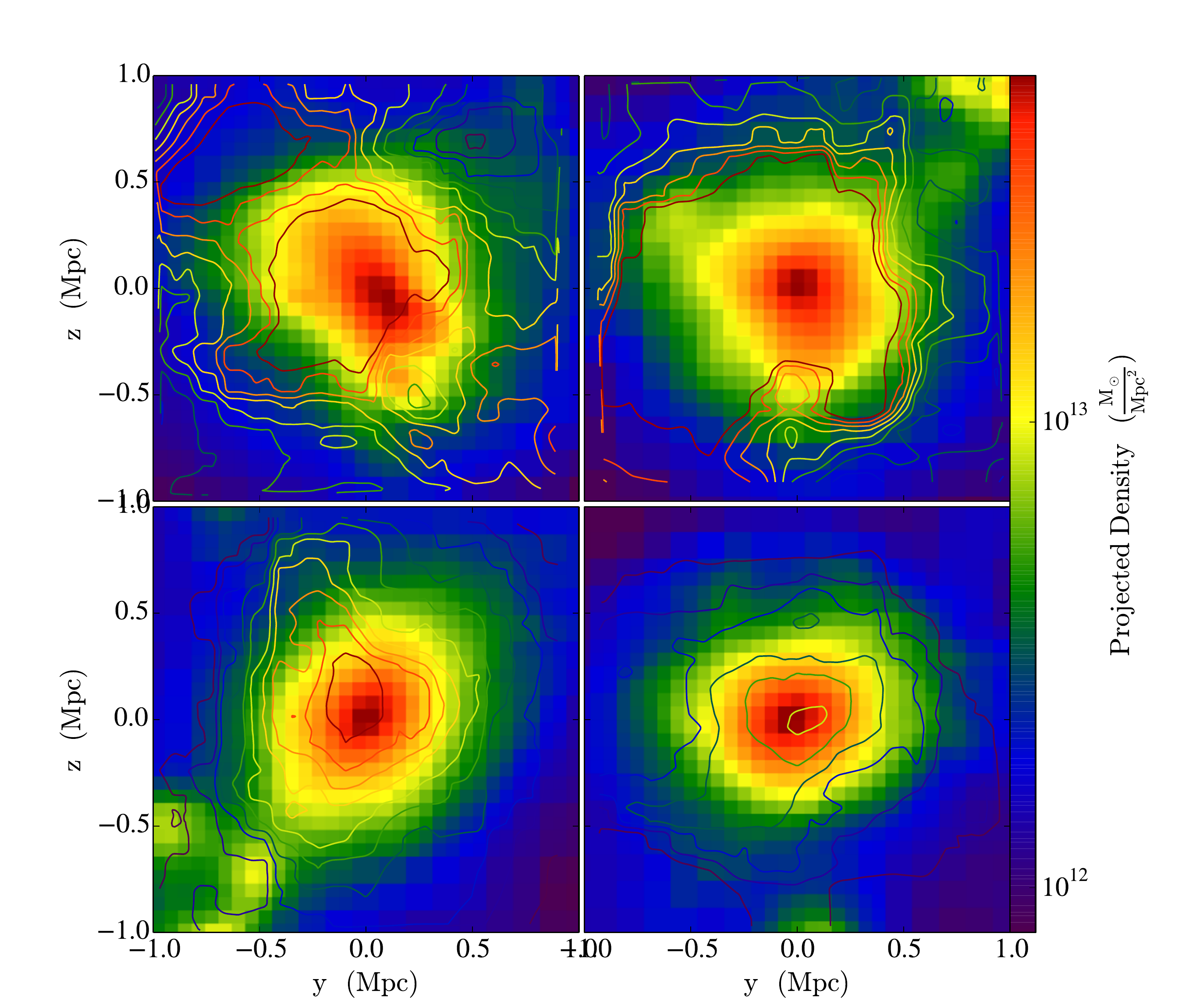}
\caption{The same four clusters as shown in Figure \ref{fig:indv_clust_LCDM}, but now from the $f(R)$ simulation shown in Figure 2 (The halos are the same because the two simulations use identical initial conditions).} 
\textcolor{white}{Some text to give caption second line. More padding text, copy as needed. More padding text, copy as needed.}
\label{fig:indv_clust_MG}
\end{minipage}
\end{figure*}
 
\subsection{Creating X-ray images}
\label{sec:ray-tracing}

To mimic W15, we need to produce X-ray profiles for our simulated haloes or clusters. This was achieved using the {\tt PHOX} software \citep{2012MNRAS.420.3545B}, as implemented in {\tt Python} \citep{2014arXiv1407.1783Z}. The software takes as primary input the output particle parameters from our hydrodynamical simulations, namely position, density, temperature and velocity. For each cluster, we input the particle information for a cube of size $2$Mpc centred on each halo. As metallicity is also required for  {\tt PHOX}, but unavailable from our simulations, we assumed a constant value of $0.3Z_{\odot}$ for all clusters, which is suitable for the outskirts of clusters \citep{2011Sci...331.1576S}.

The {\tt PHOX} software has three main steps. First, it generates a large Monte-Carlo sample of available photons (typically ten times the amount expected from an observation) in a three-dimensional volume surrounding the cluster. This is achieved by converting the given density, temperature and metallicity of each particle (or ``gas element" as described in \citealt{2011ascl.soft12004B}) into a spectrum of photons using a model for the emissivity of the intracluster medium, {\tt XSPEC} \citep{1996ASPC..101...17A} assuming a thermal {\tt APEC} model \citep{2001ApJ...556L..91S}, which is suitable for such hot, low-density, fully-ionised plasmas. Our spectral model was created with a resolution of 2000 energy bins, between $0.5$keV and $2.0$keV. For each cluster, we also created an array of different photon samples spanning a range of possible collecting areas and exposure times to facilitate the generation of realistic XMM observations in stage three of {\tt PHOX} (below).

\begin{figure}
	\includegraphics[scale=0.4]{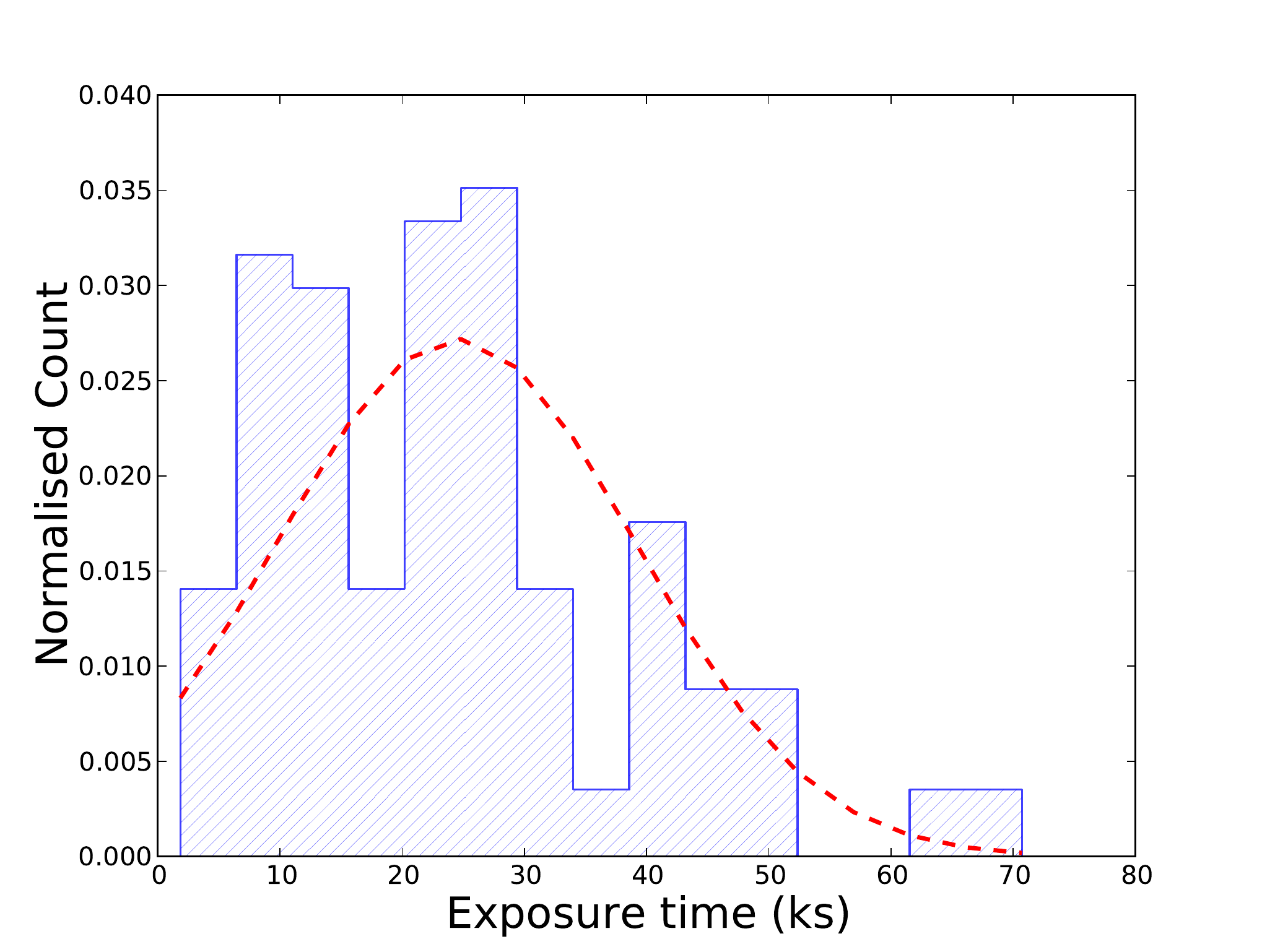}
	\captionof{figure}{The distribution of exposure times (seconds) for the real XMM cluster observations used in W15. We show the best-fit Gaussian distribution to these data.}
	\label{fig:expXCS}
\end{figure}

\begin{figure*}
\noindent\begin{minipage}{\textwidth}
    \centering
    \captionsetup{width=17cm}
	\includegraphics[scale=0.8]{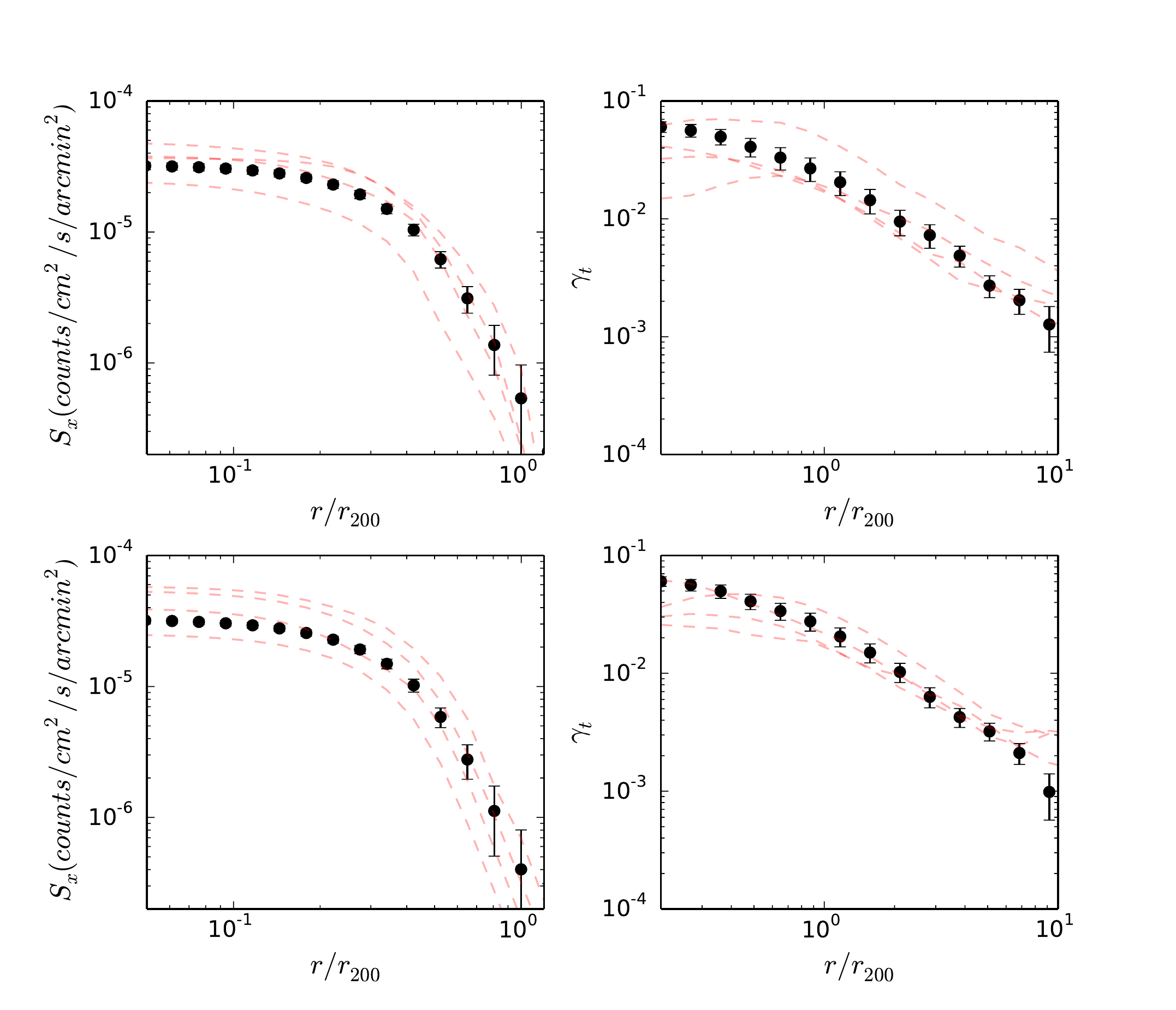}
	\captionof{figure}{(Top) The stacked X-ray surface brightness profiles (left) and lensing profiles (right) for the $\Lambda$CDM simulation (distance normalised by $r_{200}$). (Bottom) Same but for the $f(R)$ simulation.  In each case, the stacked profile is shown as black dots and is accompanied by the individual profiles of the four randomly selected clusters presented in Figures \ref{fig:indv_clust_LCDM} and \ref{fig:indv_clust_MG} (red dashed lines).}
	\label{fig:median_profiles}
\end{minipage}
\end{figure*}

%\begin{figure*}
%\noindent\begin{minipage}{\textwidth}
%    \centering
%	\includegraphics[scale=0.56]{compare_real_sim}
%	\captionof{figure}{Comparison of the real and simulated stacked profiles. The blue circles are the $T_x<2.5$ keV stacked data in W15 (which is closest to the typical masses in our simualtions), while the green crosses are for the $f(R)$ simulation and the red diamonds are the $\Lambda$CDM+GR simulation. On the left, we show the X-ray surface brightness profiles and on the right we present the lensing profiles.}
%	\label{comdata}
%\end{minipage}
%\end{figure*}

\begin{figure}
%\noindent\begin{minipage}{\textwidth}
%    \centering
%    \captionsetup{width=17cm}
	\includegraphics[scale=0.4]{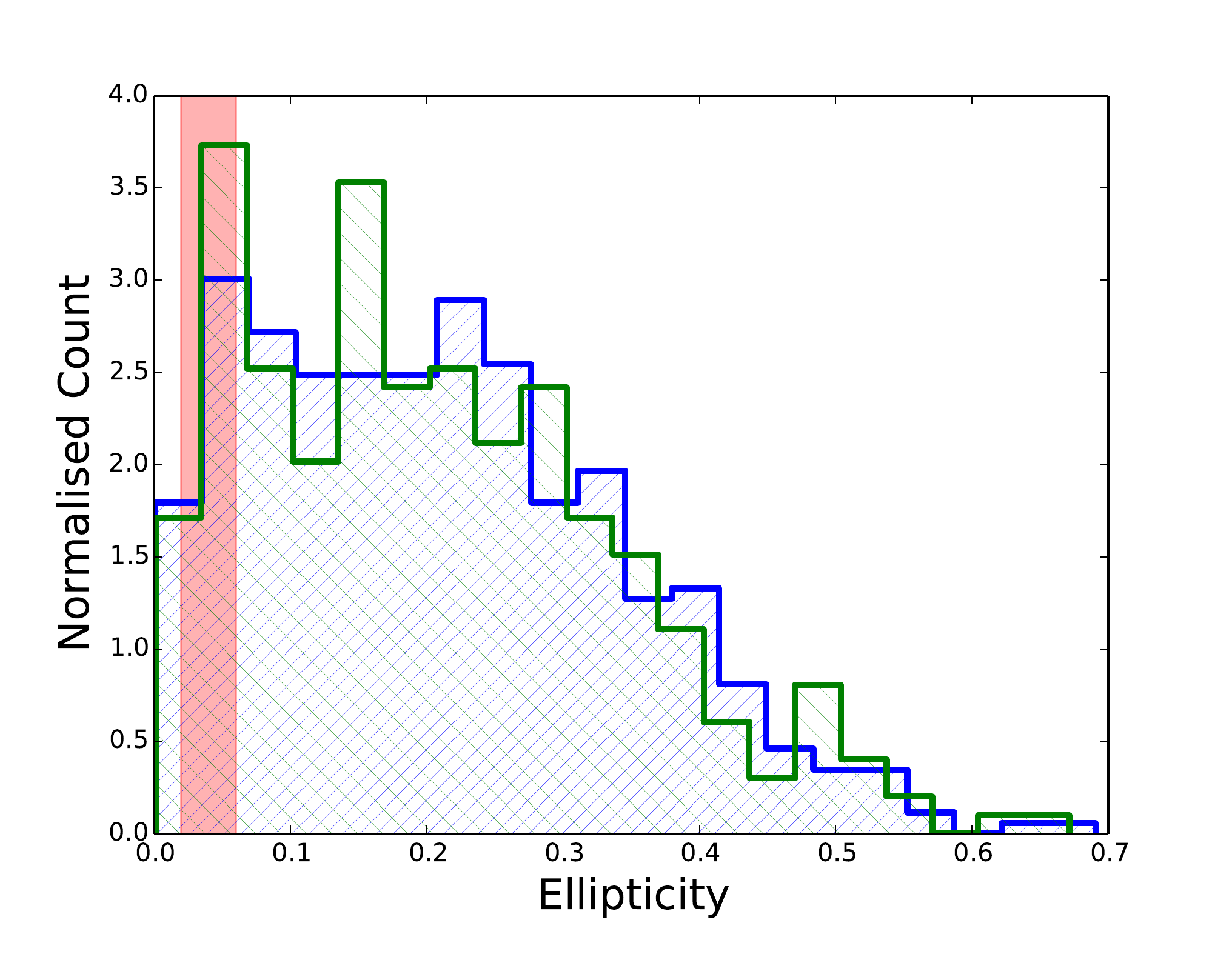}
	\captionof{figure}{The normalised distribution of ellipticities for all cluster realisations (see text) for both simulations ($\Lambda$CDM is blue and $f(R)$ is green). The (pink) shaded region shows the range of measured ellipticities after stacking the clusters.}
	\label{fig:ellip}
%\end{minipage}
\end{figure}
 
The second stage of {\tt PHOX} involves projecting the three-dimensional distribution of photons to obtain two-dimensional maps for each cluster. We select a line-of-sight for each cluster that is aligned with the z-axis in the main cosmological simulations and project the data into the plane perpendicular to this line-of-sight. This stage also corrects for doppler shifts along the line-of-sight due to the motion of the gas, and corrects for the cosmological redshift.

Next, {\tt PHOX} simulates the observing time, which is estimated using the distribution of exposure times for the real clusters in W15 (see Figure \ref{fig:expXCS}). For simplicity, we fit this distribution of exposure times with a Gaussian, giving a mean time of $24,591$ seconds and a dispersion of $12,745$ seconds.  Then, for each simulated cluster, a random exposure time is drawn from the fitted Gaussian distribution and used in {\tt PHOX} (we did not allow negative exposure times, but simply drew from the Gaussian again). The redshift for our simulated clusters is assumed to be $z=0.4$ to be consistent with the simulations. 

Finally, stage two of {\tt PHOX} corrects for the effects of absorption by gas in the Milky Way. The software uses the {\tt wabs} absorption model \citep{1983ApJ...270..119M}, which is implemented through {\tt XSPEC}, and takes the galactic column density ($N_{H}$) as an input. We assume a constant value of $10^{21} \mathrm{cm}^{-2}$ for all clusters, which is reasonable given the observed galactic column density for the W15 sample {\citep{2005A&A...440..775K}. 

The third stage of {\tt PHOX} involves adding realistic telescope effects to the simulated images. This is achieved using {\tt simx}, a convolution tool that contains the point-spread function (PSF), and detector response function, for a number of well-known X-ray telescopes. For this paper, we select the PN camera of XMM-Newton telescope. The {\tt simx} tool also adds a realistic background. 

The simulated X-ray images look cosmetically similar to real XMM data, including chip gaps and the same shape as the real XMM field-of-view. The only major difference is the lack of additional X-ray sources serendipitously detected in the outskirts of each XMM image. For comparison, a typical XCS image contains approximately ten bright serendipitously-detected point-sources per observation, with more fainter sources. We ignore this difference in our simulations as the probability of having overlapping point sources is still relatively small, and would have been corrected in the real data by excluding flux from that overlapping source when constructing cluster profiles. We also assume all our simulated clusters are observed on-axis, which is not true for the real XCS cluster sample. We investigated the effect of moving our simulated clusters off-axis using {\tt simx} and find any observed differences are significantly smaller than the simulated X-ray photon noise on these maps

\subsection{Estimating the weak lensing signal around clusters}
\label{sec:measure_shear}

We estimate the lensing shear signal around each of our simulated clusters, as our numerical simulations do not simulate the effects of gravitational lensing. We therefore calculate the expected lensing convergence, $\kappa$, as detailed in \cite{2001PhR...340..291B}, which can be approximated by 

\begin{equation}
\kappa = \frac{3 H_{0}^{2} \Omega_{m}}{2 c^{2}} \sum_{i} \Delta_{\chi i} \chi_{i} \frac{(\chi_{clust} - \chi_{i})}{\chi_{clust}} \frac{\delta_{i}}{a_{i}},
\label{eq:kappa_est}
\end{equation}

\noindent along the line-of-sight, assuming a flat universe and using the Born approximation. The summation is over comoving distance $\chi_{i}$, using bins of width $\Delta_{\chi i}$, while $H_{0}$ is the Hubble constant, $\Omega_{m}$ is the matter density, $a_{i}$ is the scale factor (in bin $i$), and $\delta_{i}$ is the overdensity in that bin.  This equation is not modified in the chameleon case \cite{2014JCAP...04..013T}.

The lensing convergence is measured in the simulations by first determining $r_{200}$ for each simulated cluster (i.e., the radius at which the average density of the halo reaches two hundred times the critical density). Then we extract a cylinder of radius $10 \times r_{200}$, centred on each cluster, but the length of the whole cosmological simulation (128 Mpc/$h$).  This cylinder is then divided into ten redshift slices (thickness $\Delta z=0.02$), and each slice was pixelated into a $100\times100$ grid.  The density in each pixel, $\rho(z_{i})$ is determined, and the overdensity in each pixel calculated as

\begin{equation}
\delta_{i} = \frac{\rho(z_{i}) - \bar{\rho}(z_{i})}{\bar{\rho}(z_{i})}.
\label{eq:overdensity}
\end{equation}

\noindent where $\bar{\rho}(z_{i})$ is the mean pixel density in each redshift slice.

$\kappa$ for each pixel is then calculated using equation (\ref{eq:kappa_est}), with the error on $\kappa$ ($\sigma_{\kappa}$) given by

\begin{equation}
\sigma_{\kappa}^{2} = \frac{3 H_{0}^{2} \Omega_{m}}{2 c^{2}} \sum_{i} \Delta_{\chi i} \chi_{i} \frac{(\chi_{clust} - \chi_{i})}{\chi_{clust}} \frac{\delta_{i}}{a_{i}} \frac{1}{\bar{n}_{i}}.
\label{eq:kappa_err}
\end{equation}
This pixellated map of convergence is converted into a shear field, from which tangential shear ($\gamma_{t}$) is then inferred, using the inversion technique given in \citet{1993ApJ...404..441K}. To make the shear measurements more realistic, we added a random shear noise component to the pixelated values behind each cluster using a distribution of shear noise values constructed directly from the galaxy source catalogue of CFHTLenS \citep{2012MNRAS.427..146H}.

\section{Testing our assumptions}

\subsection{Making stacked cluster profiles}

We follow exactly the same prescription as used in W15 to create stacked X-ray and lensing profiles for our simulated clusters. To generate the stacked X-ray surface brightness profile, we first extracted a square region of size $r_{200}$ around each individual simulated cluster and re-sampled the data, via linear interpolation, to a common grid of 500 by 500 pixels. We then stack the images, first re-scaling the overall amplitude of the images by the mean to  reduce covariances (as discussed in W15). The mean value of each pixel was then measured and binned into 19 logarithmic annuli out to $r_{200}$. As in W15, we use bootstrap re-sampling, with replacement, to estimate the errors on our stacked profiles. We created 100 mock samples from the real 103 (99) clusters available in the $\Lambda$CDM+GR ($f(R)$) simulation to replicate W15.

For the stacked lensing profile, we first estimate the tangential shear ($\gamma_{t}$) for each cluster and its noise component. The tangential shear in each pixel, around each cluster (calculated about the X-ray centroid), was binned into 19 equally spaced logarithmic bins out to a distance of $10 \times r_{vir}$.  For consistency with W15, we exclude the central $0.1 \times r_{vir}$. The shear in each bin was summed for all clusters and the mean shear measured \citep{2001astro.ph..8013M}.  This provides our stacked weak lensing profile.  We measure errors on the shear profile using the same bootstrap re-sampling method described above for the X-ray profiles.

\subsection{Testing our stacked profiles}

\begin{figure}
%\noindent\begin{minipage}{\textwidth}
%    \centering
%    \captionsetup{width=17cm}
	\includegraphics[scale=0.4]{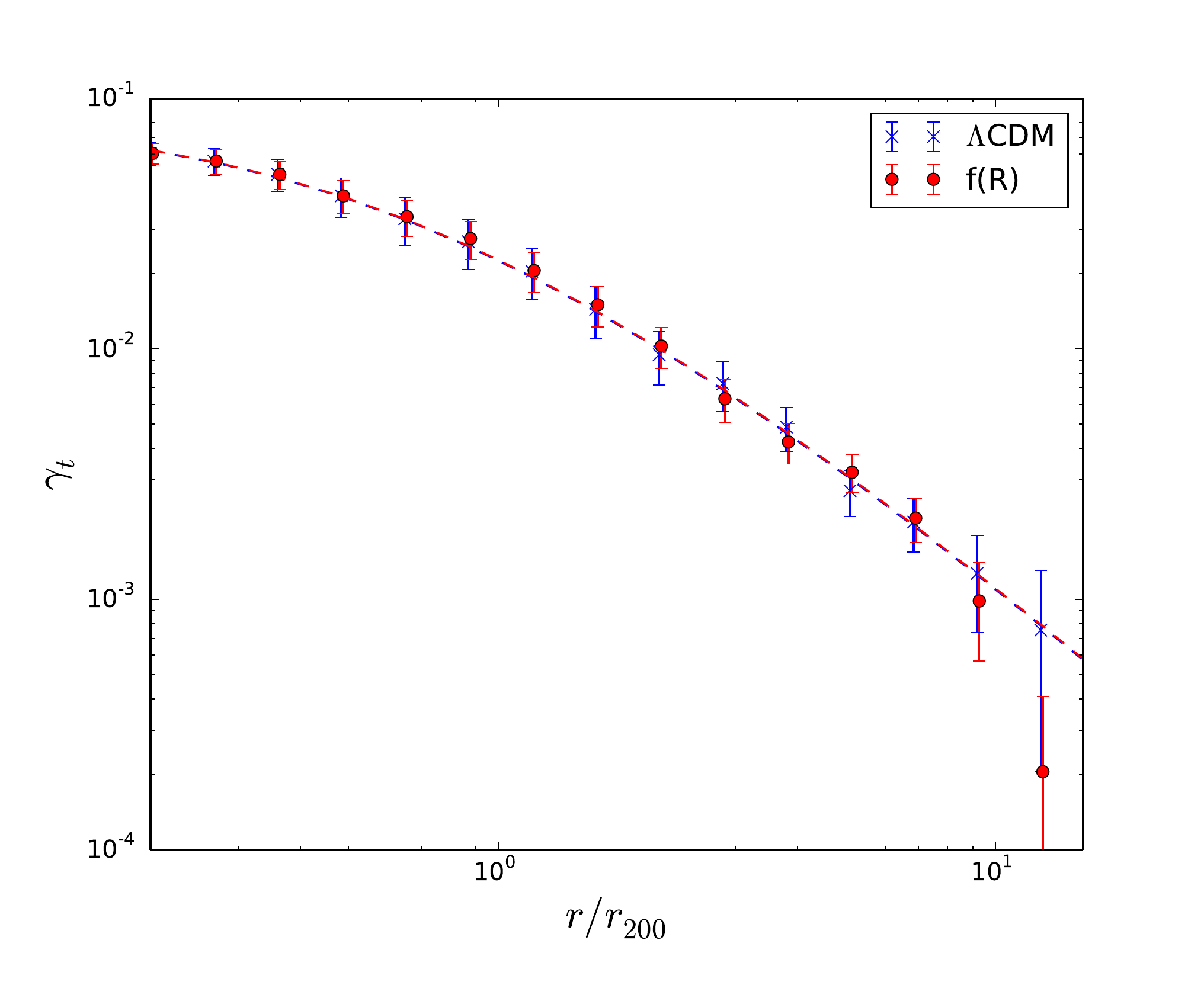}
	\captionof{figure}{We show the stacked lensing profile for the $\Lambda$CDM+GR simulation (blue crosses) and the stacked lensing profile for the $f(R)$ simulation (red  points). The dashed (blue) line is the best fit NFW profile to the $\Lambda$CDM+GR data, while the dashed (red) line is the same for the $f(R)$ profile.}
	\label{fig:NFW_compare}
%\end{minipage}
\end{figure}

In Figure \ref{fig:median_profiles}, we show the four stacked profiles (X-ray and lensing profiles for both simulations) compared to the individual profiles for the same four (randomly chosen) clusters shown in Figures \ref{fig:indv_clust_LCDM} and \ref{fig:indv_clust_MG}.

A key assumption in W15 was that stacking clusters would produce a spherically symmetric profile.  We use these simulations to explore if stacking clusters reduces possible line-of-sight projection effects which could hamper any analysis when applied to a single cluster (e.g. see Terukina et al. 2014 for a discussion of such problems for the Coma Cluster). To test this, we generate ten additional realisations per cluster, following the same methodology as given in Section \ref{sec:ray-tracing}, but now varying at random the line-of-sight direction for the projection of the three-dimensional photon distribution. We then determine the ellipticity ($\epsilon$) for each individual cluster realisation by fitting a two-dimensional ellipse to the projected surface brightness distribution (we construct an isophote where surface brightness falls to 20\% of the central pixel value). 

In Figure \ref{fig:ellip}, we show the distribution of ellipticities determined across all realisations of all our clusters in both simulations. We find a mean $\epsilon$ of $0.21\pm0.13$, which demonstrates many of our simulated clusters are non-spherical. Interestingly, we see no difference in the distribution of ellipticities between the two simulations (the mean ellipticity is also the same).

We then create ten stacked two-dimensional profiles, where each stack contained a single, different realisation of each cluster.  We fit an ellipse to each stack and computed the best-fit value of $\epsilon$. Across the ten stacks, we find $\bar{\epsilon}=0.04\pm 0.02$, which shows these stacked profiles are close to spherical (within a few percent) averaging out the ellipticities seen in the individual clusters (Figure \ref{fig:ellip}).  As our analytic model assumes spherical symmetry, knowing that our stack is also spherical gives us confidence that any constraints on $f(R)$ are not degenerate with triaxiality of the haloes.

\subsection{NFW profiles}

A key assumption made in the analysis of W15 is that the NFW profile is an appropriate model for our stacked weak lensing cluster profile. It is possible for deviations from an NFW profile to arise due to the modified dynamics during the formation of structures. In Figure \ref{fig:NFW_compare}, we show the simulated stacked weak lensing profile out to $10 \times r_{vir}$ for both the $\Lambda$CDM+GR and $f(R)$ simulation, along with the best-fit analytical NFW profiles. We used MCMC to fit the NFW parameters $c$ and $M$ (as described in W15) running the chains for 1000 time steps, removing the first 200 steps as the ``burn in'' phase. We obtain $\chi^{2}\simeq10$ (for 15 degrees of freedom) for the $f(R)$ simulation,} confirming the visual impression that the NFW model is a good representation of these lensing profiles in both simulations. 

We also find the same best-fit values of $c=7.4^{+0.64}_{-0.65}$ and $M=1.2^{+0.13}_{-0.13} \times 10^{13} M_{\odot}$ in both simulations. These values are reasonable for such dark matter haloes and consistent with other fits in the literature  (\citealt{2011JCAP...12..005P}, \citealt{2014MNRAS.440..833A}). These results confirm that an NFW profile is a good representation of the lensing profile of clusters in $f(R)$ models (as in the case of $\Lambda$CDM). This agreement is likely due to the $f(R)$ models chosen herein (F5 and F6 models), where clusters are largely screened from the modified gravity effect. We note that our test is the first time this assumption has been tested using simulated lensing profiles for $f(R)$ gravity. 

\subsection{Comparison with our analytic model}

\begin{figure}
%\noindent\begin{minipage}{\textwidth}
%    \centering
%    \captionsetup{width=17cm}
	\includegraphics[scale=0.45]{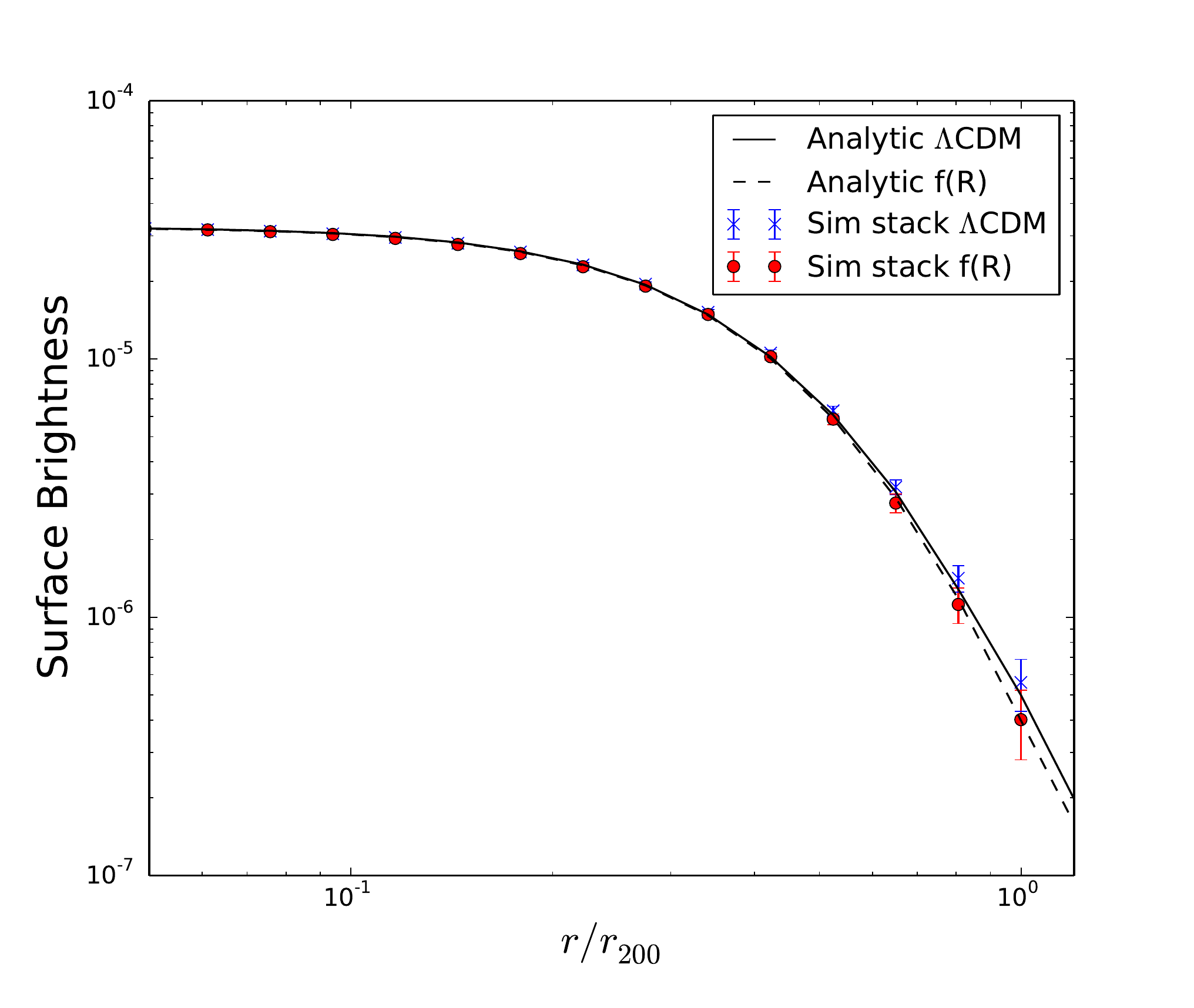}
	\captionof{figure}{The stacked X-ray profile for the $\Lambda$CDM simulation (blue crosses) and the $f(R)$ simulation in (red points). Also shown is the best-fit $\Lambda$CDM model (black line) and the best-fit $f_{\rm{R}}=10^{-5}$ line (dashed black line).}
	\label{fig:surface_brightness_test}
\end{figure}

Our simulated cluster profiles provide an ideal test for the analytical model we developed in W15 to explain the affect of chameleon gravity on the X-ray surface brightness profiles of clusters.  In Figure \ref{fig:surface_brightness_test}, we show the X-ray surface brightness profiles for both the $\Lambda$CDM+GR and $f(R)$ simulations. For comparison, we also show the expected theoretical profile using the model developed in W15 assuming GR (black line) and $f(R)$ gravity (dashed line). The latter would present itself as an additional pressure term in the hydrostatic equilibrium equation, resulting in a steeper profile in the outskirts of the cluster.  The agreement between the analytic model and the simulated model is a good validation that our analytic model can accurately describe real clusters.

In Figure \ref{fig:surface_brightness_ratio}, we show the ratio of the two simulated profiles (as the solid line) and, as discussed above, they deviate from unity in the outskirts of the cluster ($r/r_{200}>0.5$) as the gas becomes unscreened and the feels the fifth force. We also show in Figure \ref{fig:surface_brightness_ratio} our analytical prediction for this effect from W15 (assuming $f_{\rm{R}}=10^{-5}$ to be consistent with our simulation). We see the two curves agree well at small radii, while at large radii, the two are still in good agreement (always within two sigma of each other). This demonstrates that our analytical model can re-produce the overall effect of $f(R)$ gravity on the X-ray surface brightness profiles of clusters, but possibly under-estimates the amplitude of the effect at intermediate radii (with the caveat that we have not included feedback in the simulations). 

\begin{figure}
%\noindent\begin{minipage}{\textwidth}
%    \centering
	\includegraphics[scale=0.45]{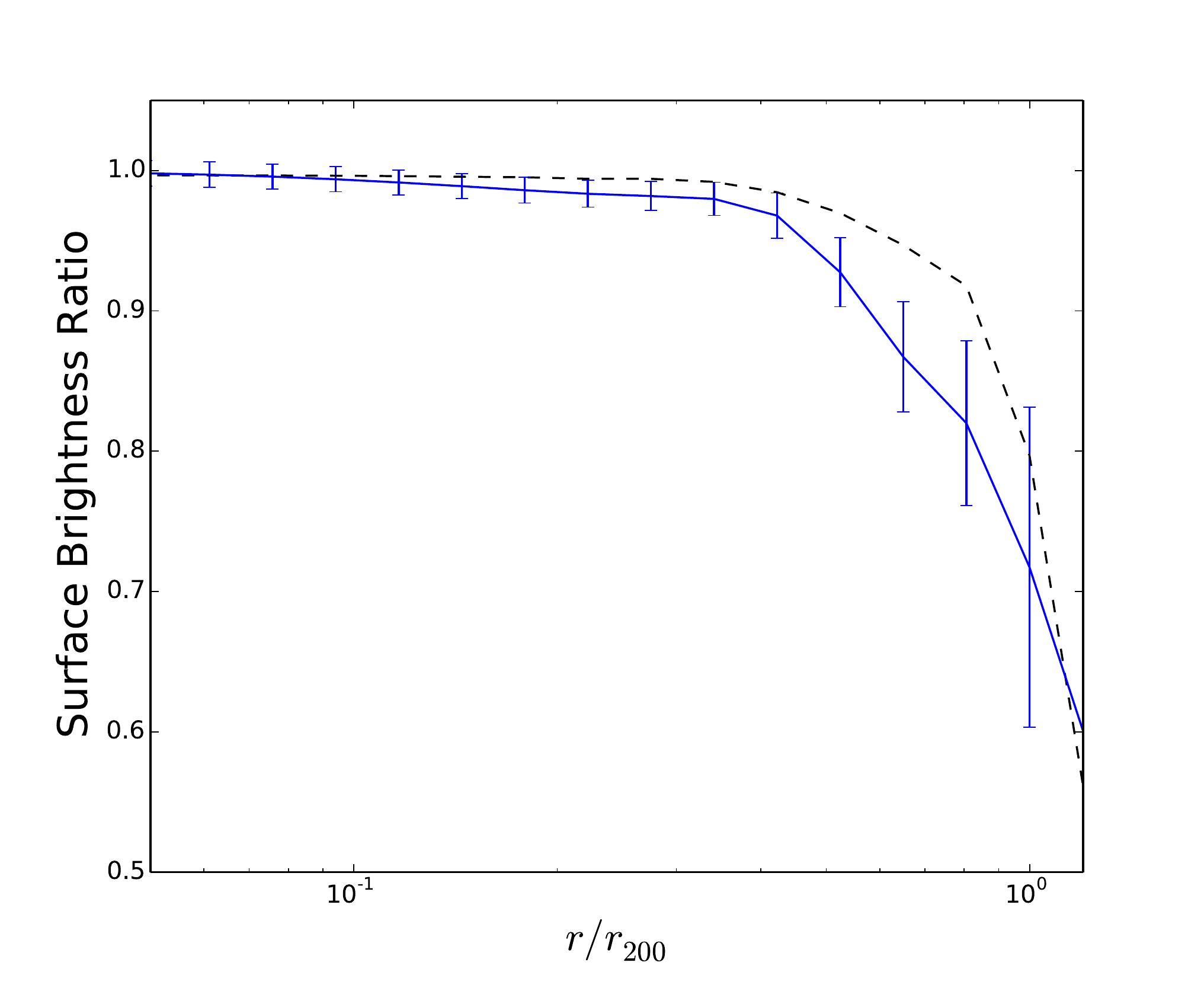}
	\captionof{figure}{The observed ratio between the two simulated stacked X-ray surface brightness profiles shown in Figure \ref{fig:surface_brightness_test} (blue line). The dashed (black) line is the same ratio but now predicted using the analytical models from Figure \ref{fig:surface_brightness_test}.}
	\label{fig:surface_brightness_ratio}
%\end{minipage}
\end{figure}

\section{Full MCMC analysis}
\label{sec:results}

\begin{figure*}
\includegraphics[width=\textwidth]{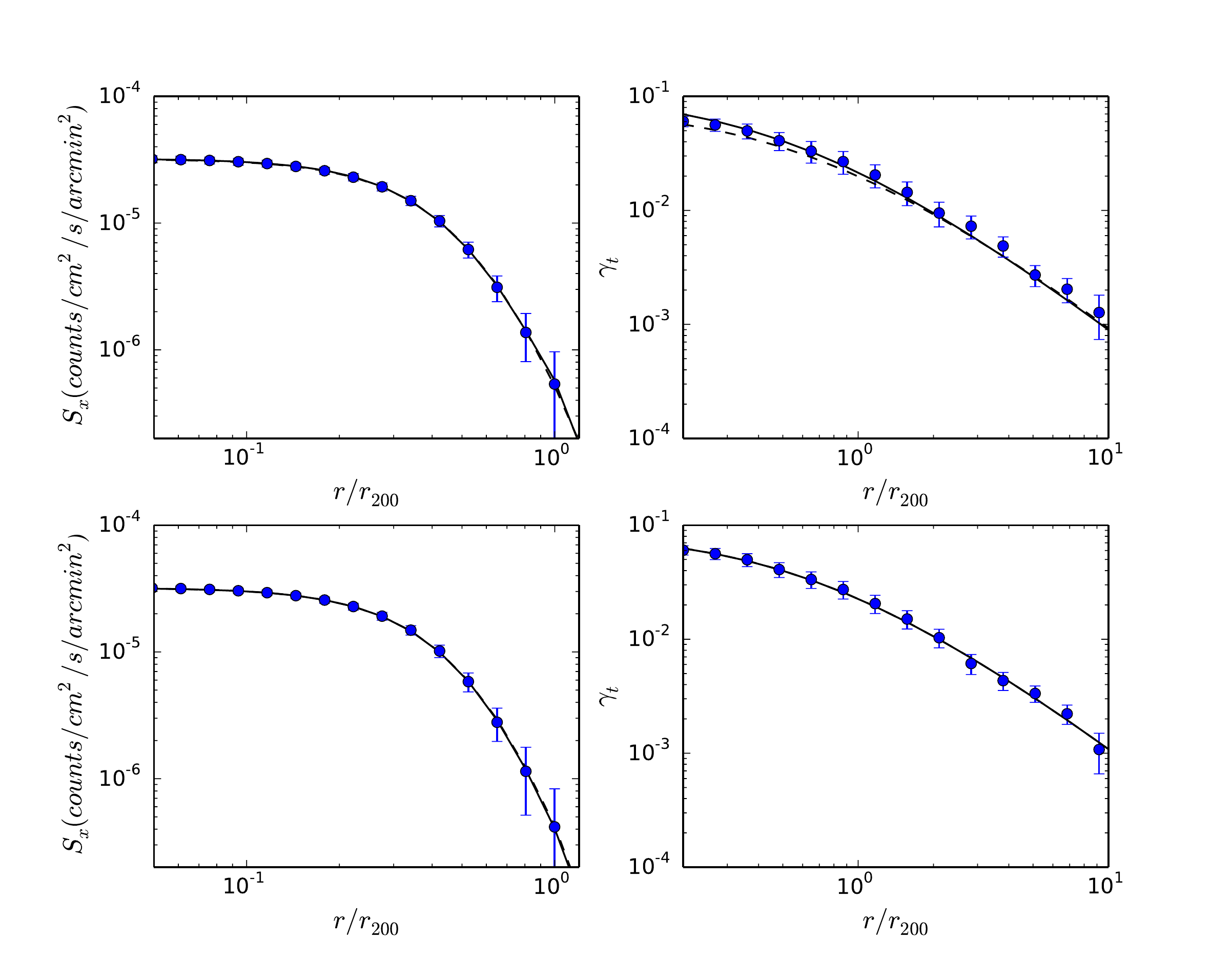}
\caption{The X-ray surface brightness (left) and weak lensing (right) profiles for the two simulations: $\Lambda$CDM+GR (top) and $f(R)$ (bottom).  For each profile, we present the 
best-fit analytical model with (dashed line), and without (solid line), the additional non-thermal pressure component. In most cases, these two fits overlap. The best-fit parameter values for the $\Lambda$CDM+GR simulation (top row), assuming no additional non-thermal pressure, are $T_{\rm{0}}=26.5$ keV, $n_{\rm{0}}=0.11 \times \rm{10^{-2} cm^{-3}} $, $b_{\rm{1}}=-2.0, r_{\rm{1}}=0.63$ Mpc, $M_{\rm{200}} = 10.0 \times \rm{10^{14}} M_{\odot}$, $c=9.0, \beta=3, \phi_{\infty}=0.7 \times 10^{-4}M_{\text{Pl}}$, which are consistent with W15 and marginalised over to get the MG parameter constraints.}
\label{fig:bestfit}
\end{figure*}

A comprehensive test of our methodology is to fit our simulated stacked profiles using the full MCMC approach described in W15, and ensure we recover the underlying cosmological parameters for our two hydrodynamical simulations (Section 2). We use the \textit{emcee} code \citep{2013PASP..125..306F} for our MCMC fitting, which implements a Metropolis-Hastings algorithm \citep{MacKayBook}. 

We provide here a brief review of the fitting technique as used in W15. We simultaneously fit our analytical model to both the stacked X-ray and lensing cluster profiles. This model is given in Equations 4 and 11 of W15, and describes both the NFW fit to the lensing profile, and the modified hydrodynamic equilibrium equation for the X-ray surface brightness profile.  Our combined model has 8 parameters, namely $c$, $M_{200}$, $n_{0}$, $b_{1}$, $r_{1}$, $T_{0}$ and the two re-scaled chameleon gravity parameters of $\beta_{2} = \beta/(1+\beta)$ and $\phi_{\infty, 2} = 1-\exp(-\phi_{\infty}/10^{-4}M_{\text{Pl}})$ (see W15 for details).

We also perform an extra fit to the profiles including an additional unknown non-thermal pressure component (e.g., \citealt{2007ApJ...668....1N}, \citealt{2009ApJ...705.1129L}) to mimic possible systematic effects on the X-ray gas at large radii (e.g. infall of cold gas onto the the cluster). As described in W15, this additional pressure component is included in the model using a parametric function for the total pressure, such that $P_{total} = g^{-1} P_{sys} = (1-g)^{-1} P_{thermal} $, where $P$ is the different pressure components, and $g$ is a function of the cluster mass and radius.

We find the best-fit model parameters using a $\chi^{2}$ statistic as described in Appendix A of W15. Our MCMC chains were run in parallel using 128 walkers with 10000 time steps (the first 2000 iterations were removed as the ``burn in'' phase). In the case of the weak lensing profile, we assume the covariance matrix is diagonal and compute it from the profile data following the technique of W15. For the X-ray surface brightness profiles, we measure the covariance matrix from the X-ray stack directly, once more following W15.

\subsection{Results}
\begin{figure*}
\centering
\begin{minipage}{0.49\textwidth}
\centering
%\exedout % first figure itself
\includegraphics[width=\textwidth, height=7.5cm]{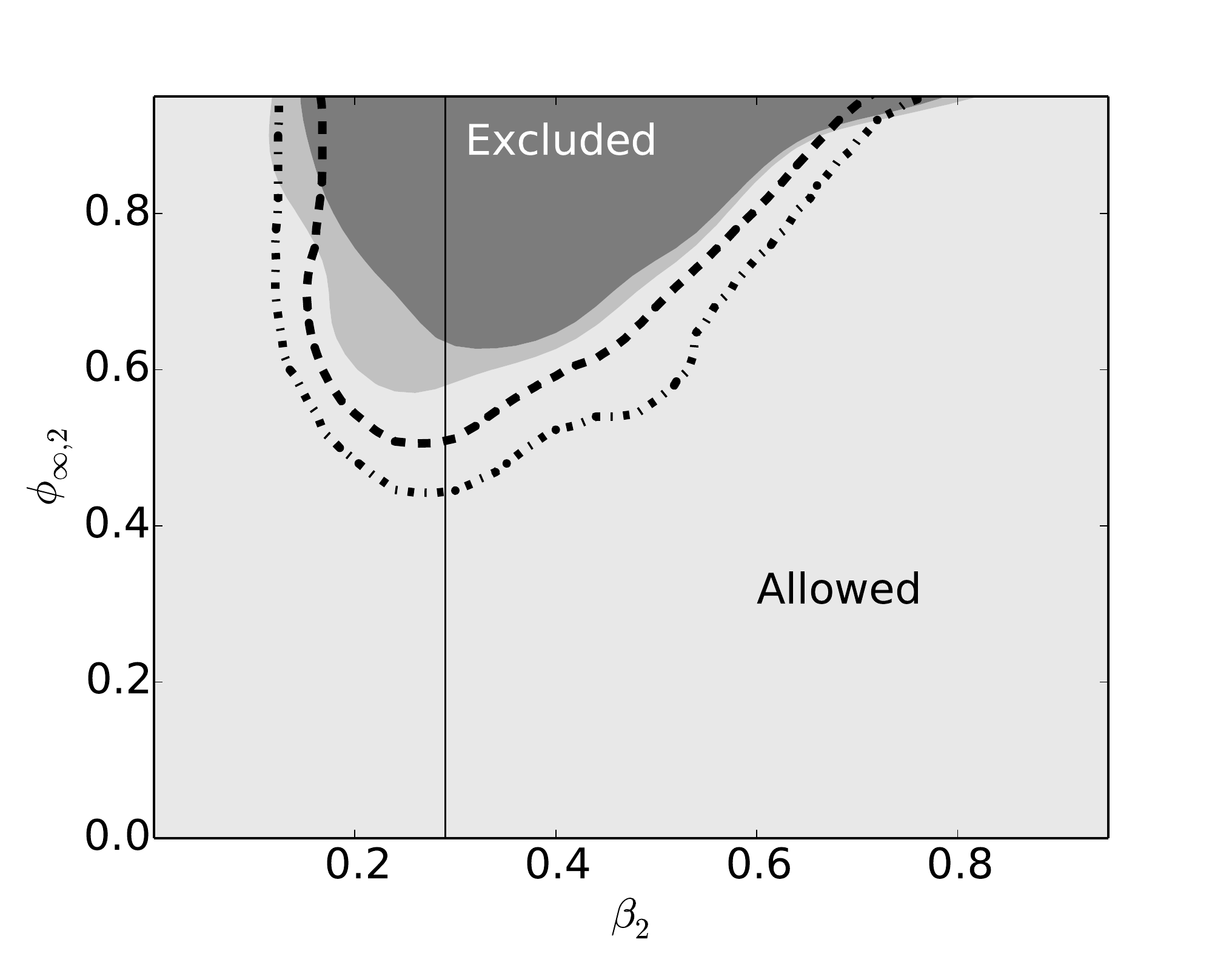}
\caption{The 95\% (light grey) and 99\% (dark grey) confidence limits for the excluded region of the combined parameter space of the two re-normalised modified gravity parameters discussed in the text. The X-ray surface brightness and lensing profiles are from the $\Lambda$CDM simulation. Also shown are confidence limits on the same parameters from W15 using the real data (dashed line is the 95 per cent, dot-dashed 99 per cent confidence) The vertical line is $\beta = \sqrt{1/6}$, showing our constraint on $f(R)$ gravity models.}
\label{fig:2D_LCDM}
\end{minipage}\hfill
\begin{minipage}{0.49\textwidth}
\centering
%\exedout % second figure itself
\includegraphics[width=\textwidth, height=7.5cm]{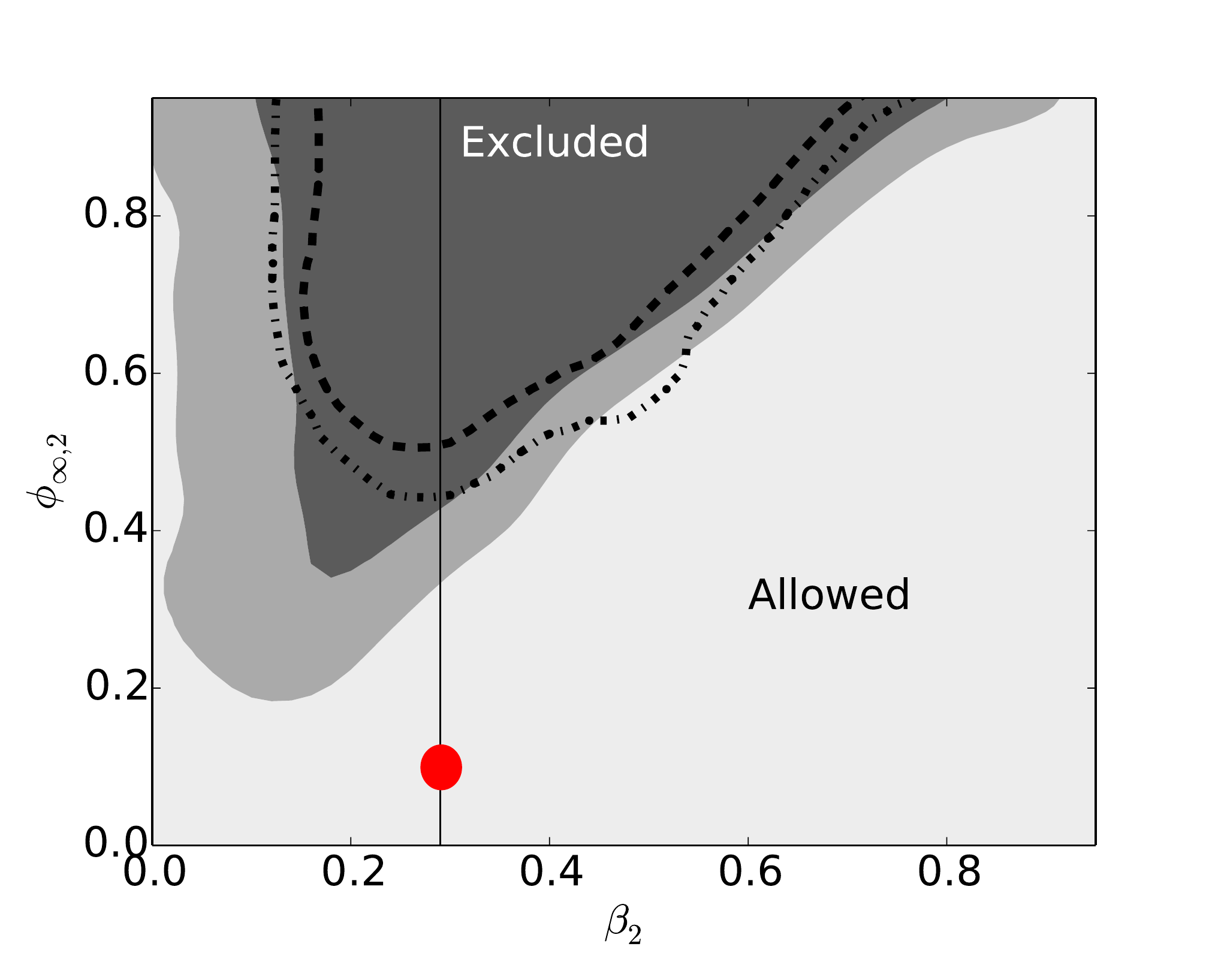}
\caption{Same as Figure \ref{fig:2D_LCDM} but for the chameleon gravity simulation. The red circle indicates the position of the fiducial model.\textcolor{white}{Some text to give caption second line. More padding text, copy as needed. More padding text, copy as needed. More padding text, copy as needed.  More padding text, copy as needed.  More padding text, copy as needed.  More padding text, copy as needed. More padding text, copy as needed. More padding text, copy as needed. More padding text, copy as needed.}}
\label{fig:2D_MG}
\end{minipage}
\end{figure*}

In Figure \ref{fig:bestfit}, we show the stacked X-ray surface brightness and weak lensing profiles from both our simulations.  We also show the best-fit model to these data with, and without, the inclusion of an additional non-thermal pressure component (as discussed above). We present the X-ray surface brightness profiles to the cluster virial radius ($r_{vir}$), while we extend the lensing profile to $10 \times r_{vir}$ (to be consistent with W15).  

% NOT SHOWING NOW
%In Appendix B, we present Figures \ref{fig:LCDM_full_contours}, 
%\ref{fig:Lcdm_Full_Contours_Nth}, \ref{fig:MG_full_contours} and 
%\ref{fig:MG_full_contours_ntp}, which show the full likelihood contours for joint %constraints between all eight parameters in our analytical model. We provide constraints %both with, and without, the additional pressure component. These joint constraints can be %compared to similar plots given in W15.

For simplicity, we focus on the two chameleon gravity parameters in our model ($\beta_{2}$ and $\phi_{\infty,2}$) and show in Figure \ref{fig:2D_LCDM} the marginalised joint constraint on these two parameters using the simulated cluster profiles from our $\Lambda$CDM+GR simulation. We also show the joint constraints obtained by W15 for these two parameters, but using the real data. 

In Figure \ref{fig:2D_MG}, we show a similar marginalised joint constraint on $\beta_{2}$ and $\phi_{\infty,2}$, but now using data from our $f(R)$ gravity simulation. We again show the constraints from W15 but with the real data. We also mark the fiducial value of these modified gravity parameters for our $f(R)$ gravity simulation. Figures \ref{fig:2D_LCDM_ntp} and \ref{fig:2D_MG_ntp} replicate these constraints, but with the additional non-thermal pressure component included (Section \ref{sec:results}). It is interesting to note that the best-fit model for the additional non-thermal pressure component in both simulations is consistent with zero, which is reassuring as neither simulation had such non-thermal physical processes added (e.g. feedback mechanisms).

These figures show that we can obtain meaningful constraints on the modified gravity parameters at a level consistent with W15. The size of the allowed regions for these joint constraints depends on the underlying simulation, and whether we include an additional pressure component or not. The most realistic constraint is given in Figure \ref{fig:2D_LCDM}, which is for $\Lambda$CDM+GR with no additional non-thermal pressure. Here our constraints are close to those found in W15, which is reassuring (assuming the true cosmological model is $\Lambda$CDM+GR). 

These joint constraints can be used to place an upper limit on $|f_{\rm{R}0}|$, which can then be compared to W15 and, in the case of the $f(R)$ simulation, the input value for that simulation. Likewise, we can place an upper limit on $f(R)$ gravity by placing a constraint on $\phi_{\infty}$ as such models are a subset of the chameleon model with $\beta = \sqrt{1/6}$ (shown as the vertical line in Figures \ref{fig:2D_LCDM}, \ref{fig:2D_MG}, \ref{fig:2D_LCDM_ntp} and \ref{fig:2D_MG_ntp}). These constraints are shown in Table \ref{table:results}, for both simulations with, and without, the extra non-thermal pressure.   We note here that the constraints recovered here are comparable to those present in W15 of $|f_{\rm{R}0}|  < 6 \times 10^{-5}$, discussed further in Section \ref{sec:discuss}.

%In our $\Lambda$CDM+GR fit without (with) a systematic error we find $\phi_{\infty} < 8.7 \times 10^{-5} M_{\text{Pl}}$ ($\phi_{\infty} < 1.1 \times 10^{-4} M_{\text{Pl}}$), and in the chameleon gravity case without (with) a systematic error $\phi_{\infty} < 4.0 \times 10^{-5} M_{\text{Pl}}$ ($\phi_{\infty} < 5.7 \times 10^{-5} M_{\text{Pl}}$).  

%Using Equation \ref{eq:f(R)}, in the $\Lambda$CDM+GR case without (with) a systematic error we place an upper limit on $f_{\rm{R}} (z=0.4) < 7.1 \times 10^{-5}$ ($f_{\rm{R}} (z=0.4) < 9.6 \times 10^{-5}$) (where $z=0.4$ is the redshift the simulations were evolved too).  Then in the chameleon gravity case without (with) additional pressure we measure $f_{\rm{R}} (z=0.4) < 3.3 \times 10^{-5}$ ($f_{\rm{R}} (z=0.4) < 4.7 \times 10^{-5}$).

\begin{table}
\centering
\begin{tabular}{l|ll}
 & \begin{tabular}[t]{@{}c@{}}Without non-\\thermal pressure\end{tabular} & \begin{tabular}[t]{@{}c@{}}With non-\\thermal pressure\end{tabular} \\ \hline 
$\Lambda$CDM - $\phi_{\infty}$ & $< 8.7 \times 10^{-5} M_{\text{Pl}}$ & $< 1.1 \times 10^{-4} M_{\text{Pl}}$ \\
$f(R)$ - $\phi_{\infty}$& $< 4.0 \times 10^{-5} M_{\text{Pl}}$ & $< 5.7 \times 10^{-5} M_{\text{Pl}}$ \\
$\Lambda$CDM - $f_{\rm{R0}}$ & $< 8.3 \times 10^{-5}$ & $< 1.1 \times 10^{-4}$ \\
$f(R)$ - $f_{\rm{R0}}$ & $< 3.8 \times 10^{-5}$ & $< 5.5 \times 10^{-5}$
\end{tabular}
\caption{Summary of constraints on modified gravity parameters from both simulations, with and without a non-thermal pressure component (95\% CL). }
\label{table:results}
\end{table}

The time-evolution of $f_{\rm{R}}(z)$ for a Hu-Sawicki model with $n=1$ (where $n$ is an additional degree of freedom of the model) follows \citep{2013MNRAS.428..743L},
\begin{equation}
f_{\rm{R}}(z) = |f_{\rm{R}0}|[(1+3\Omega_{\Lambda})/(\Omega_{\rm{M}}(1+z)^{3}+4 \Omega_{\Lambda})]^{2}.
\label{eq:f(R)_evolution}
\end{equation}
This evolution leads to a reduction in the magnitude of $f_{\rm{R}}$ by 27\% at the present day when compared with the redshift at which the simulation was placed, $z=0.4$, due to a higher background energy density at higher redshifts.  This effect has been taken into account for the values of $|f_{\rm{R}0}|$ presented in Table \ref{table:results}. When we have included fitting for a systematic error, our constraints are less stringent as the additional pressure can be degenerate with a fifth force, reducing the signal.  

\begin{figure*}
\centering
\begin{minipage}{0.49\textwidth}
\centering
%\exedout % first figure itself
\includegraphics[width=\textwidth, height=7.5cm]{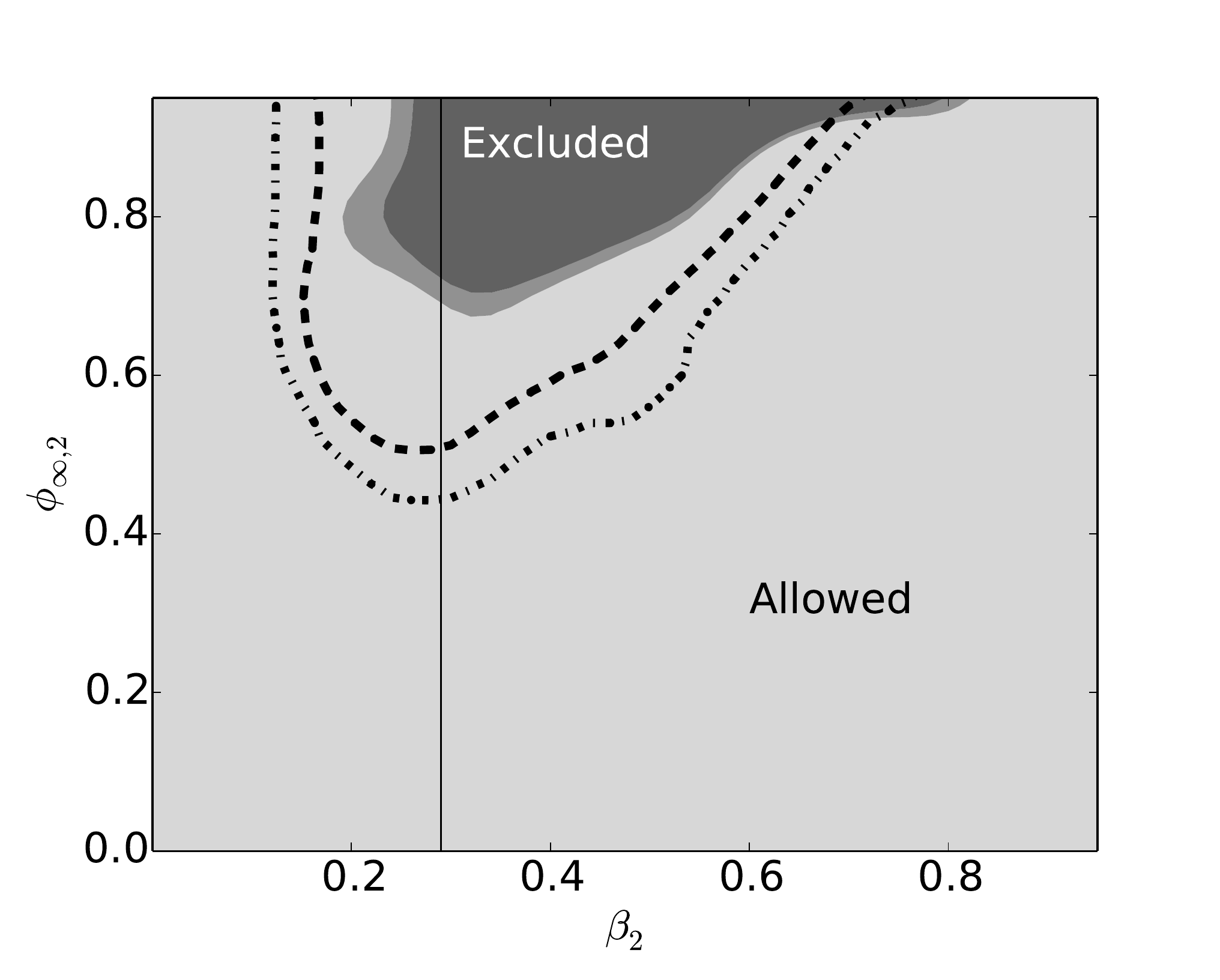}
\caption{Same as Figure \ref{fig:2D_LCDM} ($\Lambda$CDM+GR) but with a non-thermal pressure component added. \textcolor{white}{Some text to give caption second line. More padding text, copy as needed.}}
\label{fig:2D_LCDM_ntp}
\end{minipage}\hfill
\begin{minipage}{0.49\textwidth}
\centering
%\exedout % second figure itself
\includegraphics[width=\textwidth, height=7.5cm]{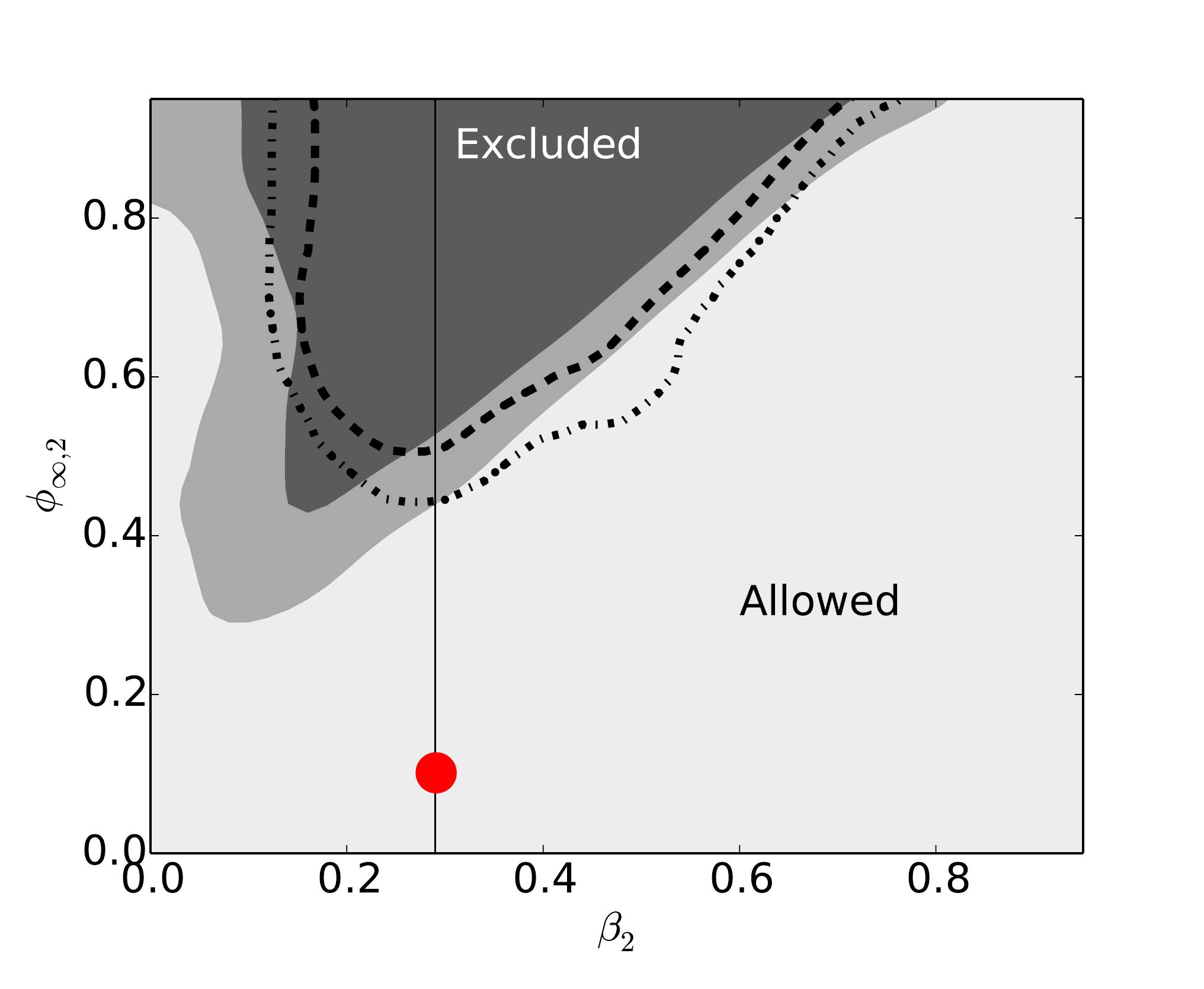}
\caption{Same as Figure 10 (chameleon gravity) but with a non-thermal pressure component added. The red circle indicates the position of the fiducial model.}
\label{fig:2D_MG_ntp}
\end{minipage}
\end{figure*}

\section{Discussion and Conclusion}
\label{sec:discuss}

In this paper, we investigate the methodology presented in W15 and test some of the assumptions made in that analysis. This is achieved using two hydrodynamical simulations; one evolved using $\Lambda$CDM+GR and the other evolved using a modified gravity component of $|f_{\rm{R}0}|=10^{-5}$.  Using these simulations, we generated realistic stacked weak lensing and X-ray surface brightness profiles.  

We use these stacked profiles to test the assumptions outlined in Section 1. We demonstrate that the stacking process created representative, spherically symmetrical profiles, thus reducing the possible bias caused by any ellipticity in an individual cluster.

%we found that the stacking process for generating high signal-to-noise cluster profiles was fair and created representative lensing and X-ray surface brightness profiles \textcolor{green}{that are in general agreement of the properties of the individual clusters that make up the stack.}  Furthermore,

We also investigated the assumption that dark matter haloes in chameleon gravity are well described by the same NFW profile as used in $\Lambda$CDM. We find no difference between the fitted NFW parameters for both our $f(R)$ and $\Lambda$CDM simulated stacked lensing profiles confirming previous studies in the literature (\citealt{2011EPJC...71.1834L},\citealt{2014MNRAS.440..833A}).% This is the first time such a test has been performed using simulated weak lensing cluster profiles (other tests have simply compared the dark matter halo density profiles and X-ray temperature profiles). 

As a complete test of the methodology, we have compared our simulations with analytical predictions used by W15. The results of this test are summarised in Figure {\ref{fig:surface_brightness_test} which shows broad agreement between the analytical and numerical (simulation) results, with the latter showing a slight deviation from $\Lambda$CDM for the same value of $|f_{\rm{R}0}|=10^{-5}$. This deviation suggests that the constraints in W15 maybe under-estimated and a correction to the analytical model could be determined using these simulations. Ideally, one would compare the simulations directly to the data, but it remains computationally-intensive to produce sufficiently large simulations for the next generation of cluster samples. For now, the analytical model remains appropriate.

In Section 4, we have replicated the full MCMC analysis from W15, but now using the simulated stacked cluster profiles instead of real data. We also include the possibility of an additional unknown non-thermal pressure component in the intracluster medium which would produce a significant systematic bias in our modelling. We present a summary of our full MCMC results in Table 1. 
%while the full likelihood contours are in Appendix C. 

For our $\Lambda$CDM+GR simulation, we find $|f_{\rm{R}0}| < 8.3 \times 10^{-5}$, which is in good agreement with the limit in W15 ($|f_{\rm{R}0}|  < 6 \times 10^{-5}$).   This validates the methodology in W15 and shows our technique can deliver competitive constraints upon the chameleon gravity model, i.e., W15 still provides one of the best constraints on $|f_{\rm{R}0}|$ on cluster (Mpc) scales. That said, the analytical model used in W15 was limited in the modelling of non-thermal pressure in the outskirts of clusters as well as the assumption of hydrostatic equilibrium. Such systematic effects are best estimated using more detailed simulations (with feedback) as discussed in this paper.

In the $f(R)$ case, we are able to recover a value of $|f_{\rm{R}0}| < 3.8 \times 10^{-5}$, which is fully consistent with the fiducial value of $|f_{\rm{R}0}|=10^{-5}$ for the simulation.  In the presence of an initial modification to gravity, there is a genuine tension between the hydrostatic and lensing profiles in the $f(R)$ simulation.  In turn this leads to less parameter space which  the model can investigate before it becomes inconsistent with one or  other  of the profiles.  This therefore rules out more area in the $\beta_{2}$ versus $\phi_{\infty,2}$ plane, leading to more powerful constraints compared to $\Lambda$CDM.

We have also constrained our profiles from both simulations including a non-thermal pressure component to account for unknown systematic uncertainties (in astrophysics or the analysis). This obviously weakens the constraints as such uncertainties are partially degenerate with any fifth force i.e., both affect the shape of the profile at large scales. Our constraints with this extra pressure term are still consistent with the fiducial model.

These simulations demonstrate that our methodology in W15 is capable of constraining chameleon gravity.  However, we note that the constraints recovered in this paper are slightly less powerful than presented in W15.  
%This is due to three effects.  Firstly, the errors determined on our simulated X-ray surface brightness profiles are larger (on large scales) than those in W15 by $\sim 50\%$.  This is due to the lower typical mass (by a factor of 2) of our simulated clusters, which produce a smaller X-ray signal (by a factor of 2) when modelled using the XCS observation times (from W15). 
%Secondly, on large scales, we recover a lower signal-to-noise lensing measurement than W15. This clearly affects our ability to constrain the modified gravity parameters as the signal we are probing is at large cluster radii (Figure 6). This is again a consequence of lower mass systems (on average) than those stacked in W15 (by a factor of 2). This effect dominates over any expected boost in signal from such lower mass clusters being less screened in chameleon gravity. To test this, we repeated our analysis but now artificially boosting the lensing signal to match that seen in W15. In this case, we measure $|f_{\rm{R}0}|  < 6.5 \times 10^{-5}$, which better reflects the constraints seen in W15.  
In W15, we split our cluster sample into two separate bins based on their X-ray temperature ($T_x<2.5$keV and $T_x>2.5$keV). We found this split in temperature (mass) provided a stronger constraint on $|f_{\rm{R}0}|$ compared to a single mass bin.  However, we are unable to replicate such binning here as the distribution of cluster temperatures and masses in the simulations is much narrower, missing the more massive ($T_x>2.5$keV) halos due to the finite volume of our simulation box, as discussed in Section \ref{sec:finding_haloes}.  We will need larger simulations to address this issue and allow us to test modified gravity effects as a function of both mass and environment (as discussed in detail in W15). Future simulations should also include more realistic feedback mechanisms.

We also conclude that we need a larger sample of clusters (with both X-ray and lensing measurements) to create higher signal-to-noise stacked profiles to further test the possibility of a fifth force. Such samples of clusters should be available soon from a number of ongoing, and future, experiments like the Dark Energy Survey (\citealt{2005astro.ph.10346T}), the KIlo Degree Survey (KIDS, \citealt{2013ExA....35...25D}), Euclid \citep{2011arXiv1110.3193L}, eROSITA and the Large Synoptic Survey Telescope (\citealt{2012arXiv1211.0310L}). These surveys should provide thousands of clusters for such tests pushing the limits on $|f_{\rm{R}0}|$ to $10^{-6}$, giving more robust constraints, which are complementary to the constraints from dwarf galaxies \citep{2013ApJ...779...39J}.

Finally, our technique can be applied to testing other theories of modified gravity that involve screening of a fifth force. Recently, we applied the same technique and data to Beyond Horndeski theories \citep{sakstein2016}, while \citet{terukina2015} used a similar methodology to test Galileon gravity using data for the Coma cluster. Such tests would require simulations similar to those presented in this paper to fully validity such methods.

%\textcolor{red}{, matching or surpassing local constraints on dwarf galaxies \citep{2013ApJ...779...39J}}. At that point, we may give up on attempting to modify GR!

\section*{Acknowledgements}
HW acknowledges the financial support of SEPNet (http://www.sepnet.ac.uk) and the ICG Portsmouth. RN, GBZ, DB, KK supported by the UK Science and Technology Facilities Council grants ST/K00090X/1 and AKR is supported by the UK Science and Technology Facilities Council grants ST/L000652/1.  KK, GBZ also acknowledge support from the European Research Council grant through 646702 (CosTesGrav).  Numerical computations were performed on the Sciama High Performance Computing (HPC) cluster which is supported by the ICG, SEPNet and the University of Portsmouth. All data from this paper is available on request from the authors.

\bibliographystyle{mn2e}
\bibliography{sim_ref}

\begin{thebibliography}{}

\bibitem[\protect\citeauthoryear{{Arnaud}}{{Arnaud}}{1996}]{1996ASPC..101...17A}
{Arnaud} K.~A.,  1996, in {Jacoby} G.~H.,  {Barnes} J.,  eds, Astronomical Data
  Analysis Software and Systems V Vol.~101 of Astronomical Society of the
  Pacific Conference Series, {XSPEC: The First Ten Years}.
p.~17

\bibitem[\protect\citeauthoryear{{Arnold}, {Puchwein} \& {Springel}}{{Arnold}
  et~al.}{2014}]{2014MNRAS.440..833A}
{Arnold} C.,  {Puchwein} E.,    {Springel} V.,  2014, MNRAS, 440, 833

\bibitem[\protect\citeauthoryear{{Bartelmann} \& {Schneider}}{{Bartelmann} \&
  {Schneider}}{2001}]{2001PhR...340..291B}
{Bartelmann} M.,  {Schneider} P.,  2001, PhysRep, 340, 291

\bibitem[\protect\citeauthoryear{{Behroozi}, {Wechsler} \& {Wu}}{{Behroozi}
  et~al.}{2013}]{2013ApJ...762..109B}
{Behroozi} P.~S.,  {Wechsler} R.~H.,    {Wu} H.-Y.,  2013, ApJ, 762, 109

\bibitem[\protect\citeauthoryear{{Biffi}, {Dolag}, {Boehringer} \&
  {Lemson}}{{Biffi} et~al.}{2011}]{2011ascl.soft12004B}
{Biffi} V.,  {Dolag} K.,  {Boehringer} H.,    {Lemson} G., , 2011, {PHOX: X-ray
  Photon Simulator}, Astrophysics Source Code Library

\bibitem[\protect\citeauthoryear{{Biffi}, {Dolag}, {B{\"o}hringer} \&
  {Lemson}}{{Biffi} et~al.}{2012}]{2012MNRAS.420.3545B}
{Biffi} V.,  {Dolag} K.,  {B{\"o}hringer} H.,    {Lemson} G.,  2012, MNRAS,
  420, 3545

\bibitem[\protect\citeauthoryear{{Bryan} et~al.,}{{Bryan}
  et~al.}{2014}]{2014ApJS..211...19B}
{Bryan} G.~L.,  et~al., 2014, APJS, 211, 19

\bibitem[\protect\citeauthoryear{Capozziello}{Capozziello}{2002}]{Capozziello:2002rd}
Capozziello S.,  2002, Int.J.Mod.Phys., D11, 483

\bibitem[\protect\citeauthoryear{{Cataneo}, {Rapetti}, {Schmidt}, {Mantz},
  {Allen}, {Applegate}, {Kelly}, {von der Linden} \& {Morris}}{{Cataneo}
  et~al.}{2015}]{2015PhRvD..92d4009C}
{Cataneo} M.,  {Rapetti} D.,  {Schmidt} F.,  {Mantz} A.~B.,  {Allen} S.~W.,
  {Applegate} D.~E.,  {Kelly} P.~L.,  {von der Linden} A.,    {Morris} R.~G.,
  2015, PRD, 92, 044009

\bibitem[\protect\citeauthoryear{{Chiba}, {Smith} \& {Erickcek}}{{Chiba}
  et~al.}{2007}]{2007PhRvD..75l4014C}
{Chiba} T.,  {Smith} T.~L.,    {Erickcek} A.~L.,  2007, Phys. Rev. D, 75,
  124014

\bibitem[\protect\citeauthoryear{{Clifton}, {Ferreira}, {Padilla} \&
  {Skordis}}{{Clifton} et~al.}{2012}]{2012PhR...513....1C}
{Clifton} T.,  {Ferreira} P.~G.,  {Padilla} A.,    {Skordis} C.,  2012, Phys.
  Rep., 513, 1

\bibitem[\protect\citeauthoryear{{de Jong}, {Verdoes Kleijn}, {Kuijken} \&
  {Valentijn}}{{de Jong} et~al.}{2013}]{2013ExA....35...25D}
{de Jong} J.~T.~A.,  {Verdoes Kleijn} G.~A.,  {Kuijken} K.~H.,    {Valentijn}
  E.~A.,  2013, Experimental Astronomy, 35, 25

\bibitem[\protect\citeauthoryear{{Foreman-Mackey}, {Hogg}, {Lang} \&
  {Goodman}}{{Foreman-Mackey} et~al.}{2013}]{2013PASP..125..306F}
{Foreman-Mackey} D.,  {Hogg} D.~W.,  {Lang} D.,    {Goodman} J.,  2013, PASP,
  125, 306

\bibitem[\protect\citeauthoryear{{Hammami}, {Llinares}, {Mota} \&
  {Winther}}{{Hammami} et~al.}{2015}]{2015MNRAS.449.3635H}
{Hammami} A.,  {Llinares} C.,  {Mota} D.~F.,    {Winther} H.~A.,  2015, MNRAS,
  449, 3635

\bibitem[\protect\citeauthoryear{{Heymans}, {Van Waerbeke}, {Miller}, {Erben}
  \& {Hildebrandt}}{{Heymans} et~al.}{2012}]{2012MNRAS.427..146H}
{Heymans} C.,  {Van Waerbeke} L.,  {Miller} L.,  {Erben} T.,    {Hildebrandt}
  H.,  2012, MNRAS, 427, 146

\bibitem[\protect\citeauthoryear{{Jain}, {Vikram} \& {Sakstein}}{{Jain}
  et~al.}{2013}]{2013ApJ...779...39J}
{Jain} B.,  {Vikram} V.,    {Sakstein} J.,  2013, ApJ, 779, 39

\bibitem[\protect\citeauthoryear{{Joyce}, {Jain}, {Khoury} \&
  {Trodden}}{{Joyce} et~al.}{2014}]{2014arXiv1407.0059J}
{Joyce} A.,  {Jain} B.,  {Khoury} J.,    {Trodden} M.,  2014, ArXiv e-prints

\bibitem[\protect\citeauthoryear{{Kaiser} \& {Squires}}{{Kaiser} \&
  {Squires}}{1993}]{1993ApJ...404..441K}
{Kaiser} N.,  {Squires} G.,  1993, APJ, 404, 441

\bibitem[\protect\citeauthoryear{{Kalberla}, {Burton}, {Hartmann}, {Arnal},
  {Bajaja}, {Morras} \& {P{\"o}ppel}}{{Kalberla}
  et~al.}{2005}]{2005A&A...440..775K}
{Kalberla} P.~M.~W.,  {Burton} W.~B.,  {Hartmann} D.,  {Arnal} E.~M.,  {Bajaja}
  E.,  {Morras} R.,    {P{\"o}ppel} W.~G.~L.,  2005, AAp, 440, 775

\bibitem[\protect\citeauthoryear{Khoury \& Weltman}{Khoury \&
  Weltman}{2004}]{Khoury:2003aq}
Khoury J.,  Weltman A.,  2004, Phys. Rev. Lett., 93, 171104

\bibitem[\protect\citeauthoryear{{Koyama}}{{Koyama}}{2015}]{2015arXiv150404623K}
{Koyama} K.,  2015, ArXiv e-prints

\bibitem[\protect\citeauthoryear{{Lau}, {Kravtsov} \& {Nagai}}{{Lau}
  et~al.}{2009}]{2009ApJ...705.1129L}
{Lau} E.~T.,  {Kravtsov} A.~V.,    {Nagai} D.,  2009, ApJ, 705, 1129

\bibitem[\protect\citeauthoryear{{Laureijs}, {Amiaux}, {Arduini},
  {Augu{\`e}res}, {Brinchmann}, {Cole}, {Cropper}, {Dabin}, {Duvet}, {Ealet} \&
  et al.}{{Laureijs} et~al.}{2011}]{2011arXiv1110.3193L}
{Laureijs} R.,  {Amiaux} J.,  {Arduini} S.,  {Augu{\`e}res} J.~.,  {Brinchmann}
  J.,  {Cole} R.,  {Cropper} M.,  {Dabin} C.,  {Duvet} L.,  {Ealet} A.,    et
  al. 2011, ArXiv e-prints

\bibitem[\protect\citeauthoryear{{Li}, {Hellwing}, {Koyama}, {Zhao}, {Jennings}
  \& {Baugh}}{{Li} et~al.}{2013}]{2013MNRAS.428..743L}
{Li} B.,  {Hellwing} W.~A.,  {Koyama} K.,  {Zhao} G.-B.,  {Jennings} E.,
  {Baugh} C.~M.,  2013, MNRAS, 428, 743

\bibitem[\protect\citeauthoryear{{Li}, {Zhao}, {Teyssier} \& {Koyama}}{{Li}
  et~al.}{2012}]{ECOSMOG}
{Li} B.,  {Zhao} G.-B.,  {Teyssier} R.,    {Koyama} K.,  2012, JCAP, 1, 51

\bibitem[\protect\citeauthoryear{{Lombriser}}{{Lombriser}}{2014}]{2014AnP...526..259L}
{Lombriser} L.,  2014, Annalen der Physik, 526, 259

\bibitem[\protect\citeauthoryear{{Lombriser}, {Koyama} \& {Li}}{{Lombriser}
  et~al.}{2014}]{2014JCAP...03..021L}
{Lombriser} L.,  {Koyama} K.,    {Li} B.,  2014, JCAP, 3, 21

\bibitem[\protect\citeauthoryear{{LSST Dark Energy Science
  Collaboration}}{{LSST Dark Energy Science
  Collaboration}}{2012}]{2012arXiv1211.0310L}
{LSST Dark Energy Science Collaboration} 2012, ArXiv e-prints

\bibitem[\protect\citeauthoryear{{Lubini}, {Tortora}, {N{\"a}f}, {Jetzer} \&
  {Capozziello}}{{Lubini} et~al.}{2011}]{2011EPJC...71.1834L}
{Lubini} M.,  {Tortora} C.,  {N{\"a}f} J.,  {Jetzer} P.,    {Capozziello} S.,
  2011, European Physical Journal C, 71, 1834

\bibitem[\protect\citeauthoryear{MacKay}{MacKay}{2003}]{MacKayBook}
MacKay D. J.~C.,  2003, {Information Theory, Inference and Learning
  Algorithms}.
Cambrdige University Press

\bibitem[\protect\citeauthoryear{{McKay} et~al.,}{{McKay}
  et~al.}{2001}]{2001astro.ph..8013M}
{McKay} T.~A.,  et~al., 2001, ArXiv e-prints

\bibitem[\protect\citeauthoryear{{Morrison} \& {McCammon}}{{Morrison} \&
  {McCammon}}{1983}]{1983ApJ...270..119M}
{Morrison} R.,  {McCammon} D.,  1983, Apj, 270, 119

\bibitem[\protect\citeauthoryear{{Nagai}, {Kravtsov} \& {Vikhlinin}}{{Nagai}
  et~al.}{2007}]{2007ApJ...668....1N}
{Nagai} D.,  {Kravtsov} A.~V.,    {Vikhlinin} A.,  2007, ApJ, 668, 1

\bibitem[\protect\citeauthoryear{{Navarro}, {Frenk} \& {White}}{{Navarro}
  et~al.}{1996}]{1996ApJ...462..563N}
{Navarro} J.~F.,  {Frenk} C.~S.,    {White} S.~D.~M.,  1996, ApJ, 462, 563

\bibitem[\protect\citeauthoryear{{Noller}, {von Braun-Bates} \&
  {Ferreira}}{{Noller} et~al.}{2014}]{2014PhRvD..89b3521N}
{Noller} J.,  {von Braun-Bates} F.,    {Ferreira} P.~G.,  2014, PRD, 89, 023521

\bibitem[\protect\citeauthoryear{{Ota} \& {Yoshida}}{{Ota} \&
  {Yoshida}}{2015}]{2015arXiv151107904O}
{Ota} N.,  {Yoshida} H.,  2015, ArXiv e-prints

\bibitem[\protect\citeauthoryear{{Oyaizu}}{{Oyaizu}}{2008}]{Oyaizu1}
{Oyaizu} H.,  2008, PRD, 78, 123523

\bibitem[\protect\citeauthoryear{{Perlmutter} et~al.,}{{Perlmutter}
  et~al.}{1999}]{1999ApJ...517..565P}
{Perlmutter} S.,  et~al., 1999, ApJ, 517, 565

\bibitem[\protect\citeauthoryear{{Planck Collaboration}, {Ade}, {Aghanim},
  {Armitage-Caplan}, {Arnaud}, {Ashdown}, {Atrio-Barandela}, {Aumont},
  {Baccigalupi}, {Banday} \& et al.}{{Planck Collaboration}
  et~al.}{2014}]{planck2013}
{Planck Collaboration} {Ade} P.~A.~R.,  {Aghanim} N.,  {Armitage-Caplan} C.,
  {Arnaud} M.,  {Ashdown} M.,  {Atrio-Barandela} F.,  {Aumont} J.,
  {Baccigalupi} C.,  {Banday} A.~J.,    et al. 2014, AAP, 571, A16

\bibitem[\protect\citeauthoryear{{Pourhasan}, {Afshordi}, {Mann} \&
  {Davis}}{{Pourhasan} et~al.}{2011}]{2011JCAP...12..005P}
{Pourhasan} R.,  {Afshordi} N.,  {Mann} R.~B.,    {Davis} A.~C.,  2011, JCAP,
  12, 005

\bibitem[\protect\citeauthoryear{{Puchwein}, {Baldi} \& {Springel}}{{Puchwein}
  et~al.}{2013}]{MGGadget}
{Puchwein} E.,  {Baldi} M.,    {Springel} V.,  2013, MNRAS, 436, 348

\bibitem[\protect\citeauthoryear{{Rasia}, {Tormen} \& {Moscardini}}{{Rasia}
  et~al.}{2004}]{2004MNRAS.351..237R}
{Rasia} E.,  {Tormen} G.,    {Moscardini} L.,  2004, MNRAS, 351, 237

\bibitem[\protect\citeauthoryear{{Riess} et~al.,}{{Riess}
  et~al.}{1998}]{1998AJ....116.1009R}
{Riess} A.~G.,  et~al., 1998, AJ, 116, 1009

\bibitem[\protect\citeauthoryear{{Romer}, {Viana}, {Liddle} \& {Mann}}{{Romer}
  et~al.}{2001}]{1999astro.ph.11499R}
{Romer} A.~K.,  {Viana} P.~T.~P.,  {Liddle} A.~R.,    {Mann} R.~G.,  2001, ApJ,
  547, 594

\bibitem[\protect\citeauthoryear{{Sakstein}, {Wilcox}, {Bacon}, {Koyama} \&
  {Nichol}}{{Sakstein} et~al.}{2016}]{sakstein2016}
{Sakstein} J.,  {Wilcox} H.,  {Bacon} D.,  {Koyama} K.,    {Nichol} R.~C.,
  2016, ArXiv e-prints

\bibitem[\protect\citeauthoryear{{Schmidt}, {Lima}, {Oyaizu} \& {Hu}}{{Schmidt}
  et~al.}{2009}]{2009PhRvD..79h3518S}
{Schmidt} F.,  {Lima} M.,  {Oyaizu} H.,    {Hu} W.,  2009, PRD, 79, 083518

\bibitem[\protect\citeauthoryear{{Simionescu}, {Allen}, {Mantz}, {Werner},
  {Takei}, {Morris}, {Fabian}, {Sanders}, {Nulsen}, {George} \&
  {Taylor}}{{Simionescu} et~al.}{2011}]{2011Sci...331.1576S}
{Simionescu} A.,  {Allen} S.~W.,  {Mantz} A.,  {Werner} N.,  {Takei} Y.,
  {Morris} R.~G.,  {Fabian} A.~C.,  {Sanders} J.~S.,  {Nulsen} P.~E.~J.,
  {George} M.~R.,    {Taylor} G.~B.,  2011, Science, 331, 1576

\bibitem[\protect\citeauthoryear{{Smith}, {Brickhouse}, {Liedahl} \&
  {Raymond}}{{Smith} et~al.}{2001}]{2001ApJ...556L..91S}
{Smith} R.~K.,  {Brickhouse} N.~S.,  {Liedahl} D.~A.,    {Raymond} J.~C.,
  2001, ApJL, 556, L91

\bibitem[\protect\citeauthoryear{{Sotiriou} \& {Faraoni}}{{Sotiriou} \&
  {Faraoni}}{2010}]{2010RvMP...82..451S}
{Sotiriou} T.~P.,  {Faraoni} V.,  2010, Reviews of Modern Physics, 82, 451

\bibitem[\protect\citeauthoryear{{Terukina}, {Lombriser}, {Yamamoto}, {Bacon},
  {Koyama} \& {Nichol}}{{Terukina} et~al.}{2014}]{2014JCAP...04..013T}
{Terukina} A.,  {Lombriser} L.,  {Yamamoto} K.,  {Bacon} D.,  {Koyama} K.,
  {Nichol} R.~C.,  2014, JCAP, 4, 13

\bibitem[\protect\citeauthoryear{{Terukina} \& {Yamamoto}}{{Terukina} \&
  {Yamamoto}}{2012}]{2012PhRvD..86j3503T}
{Terukina} A.,  {Yamamoto} K.,  2012, Phys. Rev. D, 86, 103503

\bibitem[\protect\citeauthoryear{{Terukina}, {Yamamoto}, {Okabe}, {Matsushita}
  \& {Sasaki}}{{Terukina} et~al.}{2015}]{terukina2015}
{Terukina} A.,  {Yamamoto} K.,  {Okabe} N.,  {Matsushita} K.,    {Sasaki} T.,
  2015, JCAP, 10, 064

\bibitem[\protect\citeauthoryear{{The Dark Energy Survey Collaboration}}{{The
  Dark Energy Survey Collaboration}}{2005}]{2005astro.ph.10346T}
{The Dark Energy Survey Collaboration} 2005, ArXiv e-prints

\bibitem[\protect\citeauthoryear{{Wagner}, {Schlamminger}, {Gundlach} \&
  {Adelberger}}{{Wagner} et~al.}{2012}]{2012CQGra..29r4002W}
{Wagner} T.~A.,  {Schlamminger} S.,  {Gundlach} J.~H.,    {Adelberger} E.~G.,
  2012, Classical and Quantum Gravity, 29, 184002

\bibitem[\protect\citeauthoryear{{Wilcox} et~al.,}{{Wilcox}
  et~al.}{2015}]{2015MNRAS.452.1171W}
{Wilcox} H.,  et~al., 2015, MNRAS, 452, 1171

\bibitem[\protect\citeauthoryear{{Winther}, {Schmidt}, {Barreira}, {Arnold},
  {Bose}, {Llinares}, {Baldi}, {Falck}, {Hellwing}, {Koyama}, {Li}, {Mota},
  {Puchwein}, {Smith} \& {Zhao}}{{Winther} et~al.}{2015}]{winther2015}
{Winther} H.~A.,  {Schmidt} F.,  {Barreira} A.,  {Arnold} C.,  {Bose} S.,
  {Llinares} C.,  {Baldi} M.,  {Falck} B.,  {Hellwing} W.~A.,  {Koyama} K.,
  {Li} B.,  {Mota} D.~F.,  {Puchwein} E.,  {Smith} R.~E.,    {Zhao} G.-B.,
  2015, MNRAS, 454, 4208

\bibitem[\protect\citeauthoryear{{Zhao}, {Li} \& {Koyama}}{{Zhao}
  et~al.}{2011}]{Zhao11}
{Zhao} G.-B.,  {Li} B.,    {Koyama} K.,  2011, PRD, 83, 044007

\bibitem[\protect\citeauthoryear{{ZuHone}, {Biffi}, {Hallman}, {Randall},
  {Foster} \& {Schmid}}{{ZuHone} et~al.}{2014}]{2014arXiv1407.1783Z}
{ZuHone} J.~A.,  {Biffi} V.,  {Hallman} E.~J.,  {Randall} S.~W.,  {Foster}
  A.~R.,    {Schmid} C.,  2014, ArXiv e-prints

\end{thebibliography}

\appendix
\section{Implementing $f(R)$ model in MGENZO}
\label{maths}

We present here a brief overview of the modified gravity (Hu-Sawicki) model used in this paper which has been implemented via our {\tt MGENZO} software. This software is a variant of the well-established {\tt ENZO} code and is fully described in Zhao et al. (in prep). {\tt MGENZO} has been extensively studied using several independent N-body codes including {\tt MGMLAPM} \citep{Zhao11} and {\tt ECOSMOG} \citep{ECOSMOG}. The {\tt MGENZO} code uses the same algorithm to solve for the non-linear scalar field equations as the {\tt MGMLAPM} and {\tt ECOSMOG} code. Previous results from {\tt MGENZO} have been validated against other N-body and hydro-dynamical simulations of the Hu-Sawicki model, including the code comparison work of \citet{winther2015} and \citet{2015MNRAS.449.3635H}. These papers show that all these independent codes give consistent solutions for the scalar field, as well as the power spectrum and the mass function of dark matter.

In detail, we can write the action of the $f(R)$ model as \citep{Capozziello:2002rd}

\begin{equation}
S=\int \sqrt{-g} \left [\frac{R+f(R)}{16\pi G} + \mathcal{L}_{\rm M} \right ]  {\rm d}^4 x
\label{eq:S} 
\end{equation}   

\noindent where 

\begin{equation}
\label{eq:fR} f(R)=-m^2\frac{\alpha_1(-R/m^2)^n}{\alpha_2(-R/m^2)^n+1},
\end{equation}

\noindent with $m^2=H_0^2\Omega_{\rm M}$. Under the quasi-static approximation, the equation of motion of the scalar field $\delta f_{\rm{R}}$ can be obtained as \citep{2014PhRvD..89b3521N}

\begin{equation}
\label{eq:DfR} \nabla^2  \delta f_{\rm{R}} = -\frac{a^2}{3}[\delta R(f_R)+8\pi G\delta\rho_{\rm M}],
\end{equation}

\noindent where $\delta f_{\rm{R}}=f_{\rm{R}}(R)-f_{\rm{R}}(\bar{R})$, $\delta R=R-\bar{R}$ and $\delta\rho_{\rm M} - \bar{\rho_{\rm M}}$. One can invert Equation \ref{eq:fR} to relate $R$ to $f_R$ using 

\begin{equation}
\label{eq:f_R} f_{\rm{R}}=-\frac{\alpha_1}{\alpha_2^2}\frac{n(-R/m^2)^{n-1}}{[(-R/m^2)^n+1]^2}\simeq -\frac{n\alpha_1}{\alpha_2^2}\left(\frac{m^2}{-R}\right)^{n+1},
\end{equation}  

\noindent where the approximation holds if the background cosmology is close to a $\Lambda$CDM+GR model and, in this case, one can approximate $\bar{R}$ as

\begin{equation}
\bar{R}\simeq 3 H_0^2 \left[\Omega_{M}(1+z)^{3}+4\Omega_{\Lambda} \right].
\end{equation}
At redshift $z=0$, 

\begin{equation}
\label{eq:R0bar} 
\bar{R}_0\equiv\bar{R}(z=0)\simeq 3 H_0^2 \left(1+3\Omega_{\Lambda} \right)
\end{equation}

\noindent where a flat universe is assumed. Combining Equations \ref{eq:f_R} and \ref{eq:R0bar}, one can rewrite Equation \ref{eq:f_R} in terms of $f_{\rm{R0}}$, which is the background value of $f_{\rm{R}}$ at redshift $z=0$, as

\begin{equation}
f_R\simeq f_{\rm{R0}}\left[\frac{3H_0^2(1+\Omega_{\Lambda})}{-R}  \right]^{n+1}
\end{equation}

\noindent and $\delta R$ is given explicitly as 

\begin{equation}
\label{eq:DR}
\delta R(f_{\rm{R}})=3H_0^2\left\{(1+3\Omega_{\Lambda})\left(\frac{f_{\rm{R0}}}{f_{\rm{R}}}\right)^{\frac{1}{n+1}} -\left[\Omega_M(1+z)^3+4\Omega_{\Lambda}\right] \right\}.
\end{equation}
The scalar field $f_{\rm{R}}$ can then be solved numerically by combining Equation \ref{eq:DfR} with Equation \ref{eq:DR}, given the model parameters $f_{\rm{R0}}$ and $n$ with background cosmological parameters.

The modified Poisson equation for the gravitational potential $\Phi$ can be obtained by summing the $00$ and $ii$ component of the modified Einstein equation in the Hu-Sawicki f(R) model (Hu \& Sawicki 2007), namely

\begin{equation}
\label{eq:Poisson} 
\nabla^2\Phi=\frac{16\pi G}{3}a^2\delta\rho_{\rm M}+\frac{a^2}{6}\delta R(f_{\rm{R}}).
\end{equation} 
The dynamics of the system are determined by Equations \ref{eq:DfR} and \ref{eq:Poisson}. Equation \ref{eq:DfR} is a non-linear Poisson equation, and it has to be solved numerically on a regular, or self-adaptive, grid using iteration methods \citep{Oyaizu1,Zhao11,ECOSMOG,MGGadget}. 

We chose Hu-Sawicki model as it has been implemented to several simulations and, as shown in Terukina et.al. (2014) and Wilcox et.al. (2015), is insensitive to the form of the potential of the chameleon field. Thus our constraints are applicable to a wide class of chameleon models, and contains GR as a limiting case. Thus our test also serves as a consistency check of the LCDM model.  The weak equivalence principle is not violated in this model as all matter is coupled to the scalar field universally.  However, the strong equivalence principle is violated due to the scalar field and this leads to the difference between dynamical and lensing masses.

For the $f(R)$ model, using block adaptive mesh refinement (block AMR), {\tt MGENZO} solves the non-linear Poisson equation of the scalar field (Equation \ref{eq:DfR}).  The modified Newton potential $\Phi$ can then be solved. Given $\Phi$, the hydrodynamical system for baryons and dark matter particles is numerically solved (Equations 1-4 in \citealt{2014ApJS..211...19B}).

\end{document}